\documentclass[11pt,a4paper]{article}
\usepackage{jcappub}
\usepackage{graphicx}
\usepackage{dcolumn}
\usepackage{bm}
\usepackage{float} 
\usepackage{ulem}
\usepackage{xcolor}
\usepackage{amsmath, amsfonts}
\usepackage{hyperref}
\hypersetup{colorlinks=true}
\hypersetup{pdfauthor={Name}}
\usepackage{lineno}
\usepackage{mathtools}
\usepackage{amsthm}
\usepackage{geometry}
\makeatletter
\def\ps@pprintTitle{%
   \let\@oddhead\@empty
   \let\@evenhead\@empty
   \let\@oddfoot\@empty
   \let\@evenfoot\@oddfoot
}
\def\@@author[#1]#2{\g@addto@macro\elsauthors{%
    \def\baselinestretch{1}%
    \authorsep#2\unskip\textsuperscript{
      \@for\@@affmark:=#1\do{%
       \edef\affnum{\@ifundefined{X@\@@affmark}{1}{\elsRef{\@@affmark}}}%
     \unskip\sep\affnum\let\sep=,}%
      \ifx\@fnmark\@empty\else\unskip\sep\@fnmark\let\sep=,\fi
      \ifx\@corref\@empty\else\unskip\sep\@corref\let\sep=,\fi
      }%
    \def\authorsep{\space and\space}%
    \global\let\sep\@empty\global\let\@corref\@empty
    \global\let\@fnmark\@empty}%
    \@eadauthor={#2}%
    \g@addto@macro\useauthors{#2; }%
}
\makeatother

\title{Searching for Dark Matter Annihilation with IceCube and P-ONE}

\author[a]{K. Desai,}
\author[a,b]{R. Li}
\author[a,c]{and S. Meighen-Berger}
\affiliation[a]{Ludwig Maximilian University, Geschwister-Scholl-Platz 1, 80539 M\"unchen, Germany}
\affiliation[b]{Technische Universit\"at M\"unchen, James-Franck-Stra{\ss}e, 85748, Garching, Germany}
\affiliation[c]{School of Physics, The University of Melbourne, Victoria 3010, Australia}
\emailAdd{desai.kruteesh@campus.lmu.de}
\emailAdd{ruohan.li@tum.de}
\emailAdd{stephan.meighenberger@unimelb.edu.au}

\date{\today}

\abstract{
   We present a new search for weakly interacting massive particles utilizing ten years of public IceCube data, setting more stringent bounds than previous IceCube analysis on massive dark matter to neutrino annihilation. We also predict the future potential of the new neutrino observatory, P-ONE, showing that it may even exceed the sensitivities of Fermi-LAT gamma-ray searches by about 1-2 orders of magnitude in 1-10 TeV regions. This analysis considers the diffuse dark matter self-annihilation to neutrinos via direct and indirect channels, from the galactic dark matter halo and extra-galactic sources. We also predict that P-ONE will be capable of pushing these bounds further than IceCube, even reaching the thermal relic abundance utilizing a galactic center search for extended run-time.
}
\begin{document}
\maketitle

\section{Introduction}\label{sec:Intro}
New neutrino telescopes under construction, such as P-ONE\cite{PONE}, KM3NeT\cite{KM3NeT}, GVD\cite{GVD}, and IceCube-Gen2\cite{Ice_Gen2}, will improve the global sky coverage. This will lead to an increased detection sensitivity from both diffuse and point-like signals. With these new detectors in mind, we estimate neutrino telescope sensitivities to dark matter (DM) annihilation and set constraints using ten years of public IceCube data. Already, multiple dark matter studies using IceCube\cite{Ice_gen1} and/or ANTARES\cite{ANTARES_dete} have been performed, such as the search for dark matter from the Galactic Center \cite{GC_Hesse, 44_ice_GC_GH}, earth\cite{earth_center}, Sun\cite{sun_search_ice, GC_sun}, multi-messenger searches \cite{Murase2012}, and diffuse searches for decaying or interacting dark matter \cite{MostRencentIceCube, j_b_diffuseDM, Esmaili:2014rma, Chianese:2016opp, Chianese:2016kpu, Bhattacharya:2019ucd}.

Here we consider Majorana weakly interacting massive particles (WIMPs) as dark matter candidates. In the case of dark matter self-annihilation to neutrino pairs, the spectrum would possess a distinct shape: a peak at the DM rest mass. This feature differs fundamentally from the measured diffuse astrophysical neutrino power-law spectrum\cite{44_ice_GC_GH,Ice_dm_GH}, as well as the atmospheric background. Due to this, we perform energy-binned likelihood analyses, searching for these exotic energy distributions.

In standard WIMP freeze-out scenarios, DM particles are assumed to be in thermal equilibrium, with the expanding universe, the dark sector decouples due to the expansion rate increasing greater than the interaction rate, ceasing DM production and self-interaction leading to the relic density \cite{freez_1,freez_2,freez_3}. In this scenario, to account for current DM observations, the thermally averaged annihilation cross section would then be $\left \langle\sigma \nu\right \rangle$ = $3\times 10^{-26}$ $\text{cm}^3\text{s}^{-1}$ \cite{freez_1}. Thus, should DM not be found, the searches aim to push the constraints on the cross-section below this value, excluding these WIMP scenarios. Note that for DM with mass, $m_{\chi}>10 \text{ GeV}$ the thermal relic abundance constraint on the cross-section shows minor mass dependence\cite{thermal_relic_var}.

We follow \cite{thermal_relic_var} when calculating the direct neutrino annihilation channel. In this case, the WIMP spin plays an insignificant role and hence can be neglected\cite{Ice_dm_GH}. For indirect channels, we rely on the simulations performed in \cite{Cook_book} for their neutrino production spectrum. The data released in \cite{Cook_book} include various intermediate particle states, such as $W$s, $b$s, and $\tau$ s, and their consequent decay products, e.g. neutrinos, electrons, etc. In \autoref{sec:modeling} and \autoref{sec:Background}, we give a thorough discussion of the signal and background modeling for P-ONE\cite{PONE} and IceCube\cite{Ice_data}. The final limits on the thermally averaged cross-section are shown in the \autoref{sec:Limits}.

\section{Signal Modeling}\label{sec:modeling}
This section briefly describes the calculation procedure for differential neutrino fluxes produced via DM annihilation measured by a neutrino telescope on Earth. The first subsection is dedicated to neutrino pair production from the galactic DM halo. In the second subsection, we discuss the modeling of an extra-galactic signal. We adopt the signal flux simulation described in \cite{carlos, MostRencentIceCube, Lopez_, watson}.
\\
For this analysis, we consider DM self-annihilation to neutrino pairs, via a generic mass resonance, or W-boson and $\tau$-leptons. Therefore, we developed a simulation software in which we can include various types of DM decay and annihilation channels(for example annihilation into W-bosons, b-quarks, etc.\footnote{\url{https://github.com/MeighenBergerS/pone_dm/releases/tag/v1.0.0}}). We compare the direct annihilation process as well as indirect channels via W-bosons and $\tau$-leptons in \autoref{sec:Limits} with previous analyses.

\subsection{Galactic Contribution}
In the case of direct annihilation, the galactic dark matter halo contributes trivially due to the negligible redshift. This means the flux spectrum arriving at Earth is almost identical to the spectrum at the production sites, given by \autoref{eq:galactic}, assuming an equal neutrino flavor decomposition 1:1:1 after long-distance propagation
\begin{equation}
     \frac{d\Phi_{galactic}}{d E} = \frac{1}{4\pi}\frac{\left \langle\sigma\nu\right \rangle}{\kappa m_{\chi}^2}\frac{1}{3}\frac{dN_{\nu}}{dE_{\nu}} J(\Omega).
    \label{eq:galactic}
\end{equation}

$\left \langle\sigma \nu\right \rangle$ is the thermally averaged self-annihilation cross-section. We have used $\kappa = 2$ for Majorana DM with mass $m_\chi$ for the DM mass. The factor of 1/3 is due to equal distribution among neutrino flavors. $dN/dE$ is the number spectrum of the neutrinos. \autoref{eq:dn_gal} depicts the number spectrum of neutrinos produced via DM direct annihilation to neutrinos that have a typical delta peak
\begin{equation}
    \frac{dN_{\nu}}{dE_{\nu}} = 2\delta \left(1-\frac{E}{m_{\chi}}\right)\frac{m_{\chi}}{E^2}.
    \label{eq:dn_gal}
\end{equation}

The spectrum shape varies depending on the annihilation channels. DM annihilation to W-boson pairs, b-quark pairs, and $\tau$-lepton pairs are conventional choices. These heavy Standard Model (SM) particles can again decay into stable SM particles such as electrons, gamma-rays, or neutrinos. Modeling of the branching ratios is required to include different channels. In \cite{Cook_book}, thorough modeling has been performed using PYTHIA\cite{Pythia} and HERWIG\cite{Herwig} for the DM mass range from 100 GeV to 100 TeV.

For the direct annihilation to neutrinos, we neglect the Electroweak (EW) corrections. These would generate tails for the annihilation spectrum, increasing the yield of low-energy neutrinos and broadening the peak. As shown in \cite{EWcorrection, MostRencentIceCube, Bauer:2020jay, Ciafaloni:2010ti, carlos, ice_dm_mass_reso}, the broadening of the peak is less than 10\%, which is well below the typical energy resolution for track-like events in IceCube \cite{ice_recon}. These corrections would be significant in the case of cascade-like events, where the energy reconstruction is better. Of note, is that these corrections are relevant when considering other final states, such as $\gamma$, $e^+$, and $\bar{p}$, since these corrections induce a significant spectrum \cite{Ciafaloni:2010ti}.
Contrary to the direct production channel, for the indirect channels, the EW corrections have been included in the spectrum from PPPC4DM \cite{Cook_book}. In \autoref{app:spectrum}, we show the spectra for different annihilation channels. 
\\
The $J$ in \autoref{eq:galactic}, is the 'J-factor', which stands for the three-dimensional integration of the host galaxy's DM density $\rho_{\chi}$ over a solid angle $\Omega$, within the line of sight $l.o.s.$
\begin{equation}
    J(\Omega) = \int d\Omega \int_{l.o.s.} \rho_{\chi}^2 (x) dx.
    \label{eq:J}
\end{equation}

In the case of P-ONE, we use three distinct regions. The J-factor values are listed in \autoref{tab:J}. In \autoref{tab:J} we give three J-factors: s-wave, p-wave, and d-wave. The s-wave corresponds to the thermally averaged cross section independent of velocity, p-wave to  [$\left \langle\sigma\nu\right \rangle$ $\propto$ $(v/c)^2]$ and d-wave $[\left \langle\sigma\nu\right \rangle$ $\propto$ $(v/c)^4]$ each denoted as $J_{s}$, $J_{p}$, and $J_{d}$ respectively. The p-wave and d-wave contributions are suppressed due to the typical Virial velocity of 100 km/s, hence, they can be neglected.The sensitivity region of the sky for different neutrino detectors affects the value of the integrated J-factor, as presented in \cite{carlos,MostRencentIceCube}. 

We modify the J-factor in IceCube's case, splitting it into northern (up-going) and southern (down-going) sky. We then focus on the up-going events, suppressing most of the atmospheric muon background. While this effectively removes most of this background, it also removes the galactic center from the field of view. Depending on the core or cuspy nature of the dark matter distribution, this can make a significant difference. Here, we use a cusped Navarro–Frenk–White (NFW) profile \cite{NFW}, which leads to a J-factor $\sim 5$ times smaller than the All-sky one shown in \autoref{tab:J} and \cite{carlos,MostRencentIceCube}. This results in a sensitivity drop of $\sim 2$. 
Note that using a different profile, such as an Einasto \cite{1989A&A...223...89E} or Burkert \cite{Burkert:1995yz}, leads to an approximately three times more or less stringent constraint, respectively. This has been widely studied in various previous works\cite{Ice_dm_GH, KM3NET_bergmes,46_ice_Gcluster}.

\begin{table}[H]
    \centering
    \begin{tabular}{c|c|c|c}
        Experiment & $J_s/10^{23}$ & $J_p/10^{17}$ & $J_d/10^{11}$ \\
        \hline
         All-sky & 2.3 & 2.2 & 3.6 \\
         Northern-sky & 1.8 & 1.7 & 2.7 \\
         Southern-sky & 0.5 & 0.5 & 0.9 \\
        \hline
         P-ONE & & &   \\
         $cos(\theta) = [-1, -0.5]$ & 0.87 & 0.85 & 1.4 \\
        $cos(\theta) = [-0.5, 0.5]$ & 1.2 & 1.2 & 2.0 \\
         $cos(\theta) = [0.5, 1]$ & 0.13 & 0.12 & 0.18
    \end{tabular}
    \caption{J-factors calculated using the NFW density profile\cite{NFW}, for IceCube (All-sky) and P-ONE\cite{carlos}. The units of these J-factors are $\text{GeV}^2 \text{cm}^{-5} \text{sr}$. $\theta$ is the zenith angle. The columns are J-factors for the s-wave, p-wave, and d-wave contributions. For our analysis, we neglect the p-wave and d-wave contributions, due to their strong suppression.}
    \label{tab:J}
\end{table}

For the calculation of these J-factors, we assume the sun is located at a distance of $R_0 = 8.127\,\text{kpc}$ from the galactic center (GC), as determined by \cite{GC_sun}.
\\
The dark matter density used to calculate the J-factor in \autoref{eq:J} is parameterized as
\begin{equation}\label{eq:rho_s}
    \rho_{\chi}(r) = \rho_{s}\frac{2^{3-\gamma}}{(r/r_{s})^{\gamma}(1+r/r_{s})^{3-\gamma}}.
\end{equation}

\begin{figure}
    \centering
    \includegraphics[width=8.6cm]{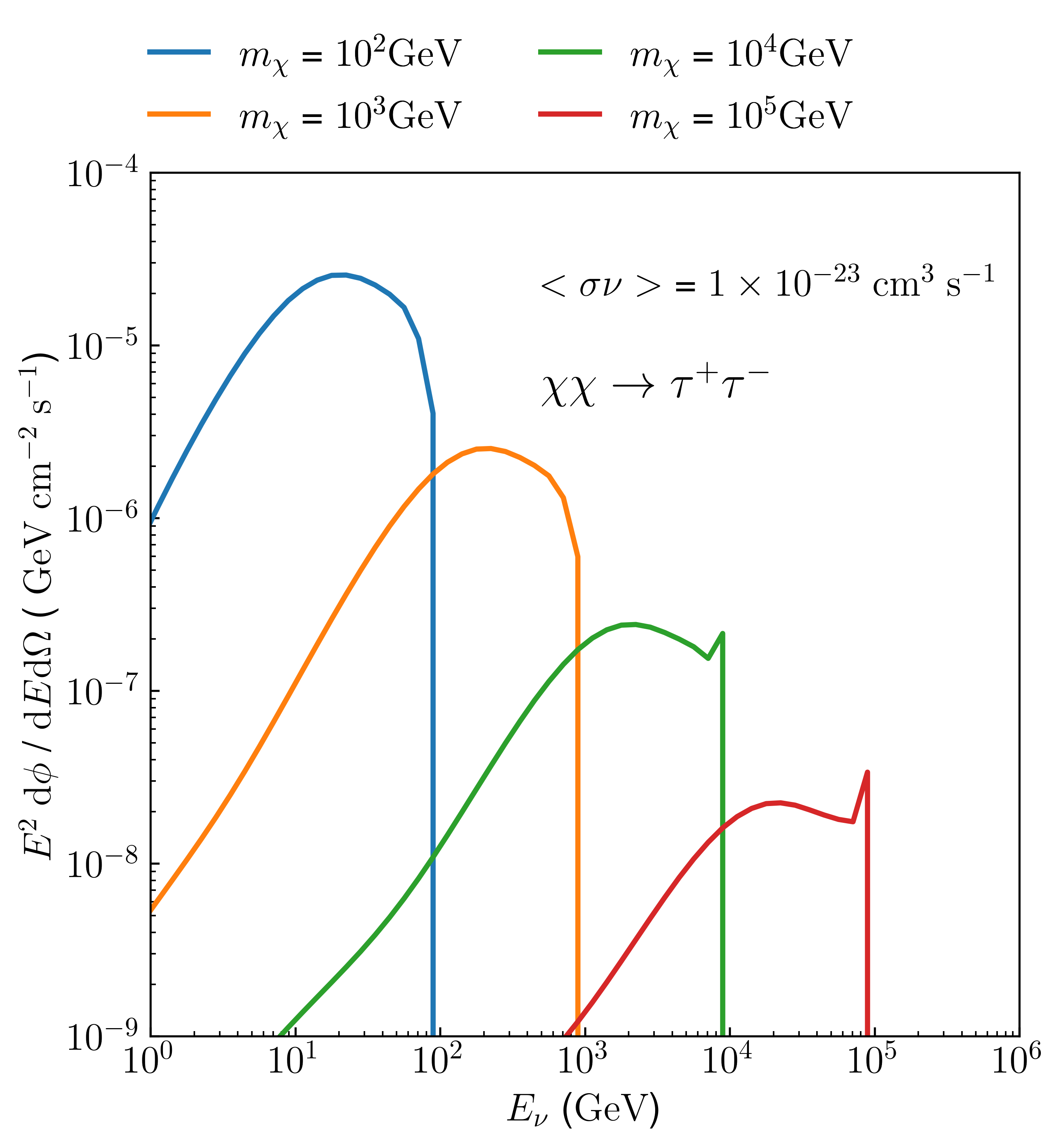}
    \caption{Differential neutrino fluxes for the galactic halo for various DM masses between 100 GeV and 100 TeV. Here $\left \langle\sigma\nu\right \rangle=1\times 10^{-23}$ $\text{cm}^{3}\text{s}^{-1}$, and the J-factors for P-ONE, see \autoref{tab:J}, are used. The differential fluxes are calculated with the spectrum data from PPPC4DM\cite{Cook_book} for DM annihilation to neutrino via the indirect $\tau$-channel.}
        \label{fig:flux}
\end{figure}

Here, $r$ is the distance from GC. We use the best-fit values from \cite{DM_para}, with a local density of $\rho_0=0.4 \, \text{GeV} \text{cm}^{-3}$, a slope parameter $\gamma=1.2$, and a scale density $\rho_s$ at scale radius $r_s=20 \, \text{kpc}$. Inverting \autoref{eq:rho_s} and setting $\rho_{\chi}(R_{0}) \equiv \rho_{0}$ we obtain $\rho_{s}$. In \autoref{fig:flux} we show the differential neutrino flux at the earth, produced by DM annihilation with various masses from the galactic halo. There, we adopt $\left \langle\sigma\nu\right \rangle$ = $1\times10^{-23}$ $\text{cm}^{3}\text{s}^{-1}$ and consider the neutrino production via DM annihilation to $\tau$-pairs. The production spectra are taken from PPPC4DM\cite{Cook_book}.

\subsection{Extra-Galactic Contribution}

\begin{figure}
        \centering
        \includegraphics[width=8.6cm]{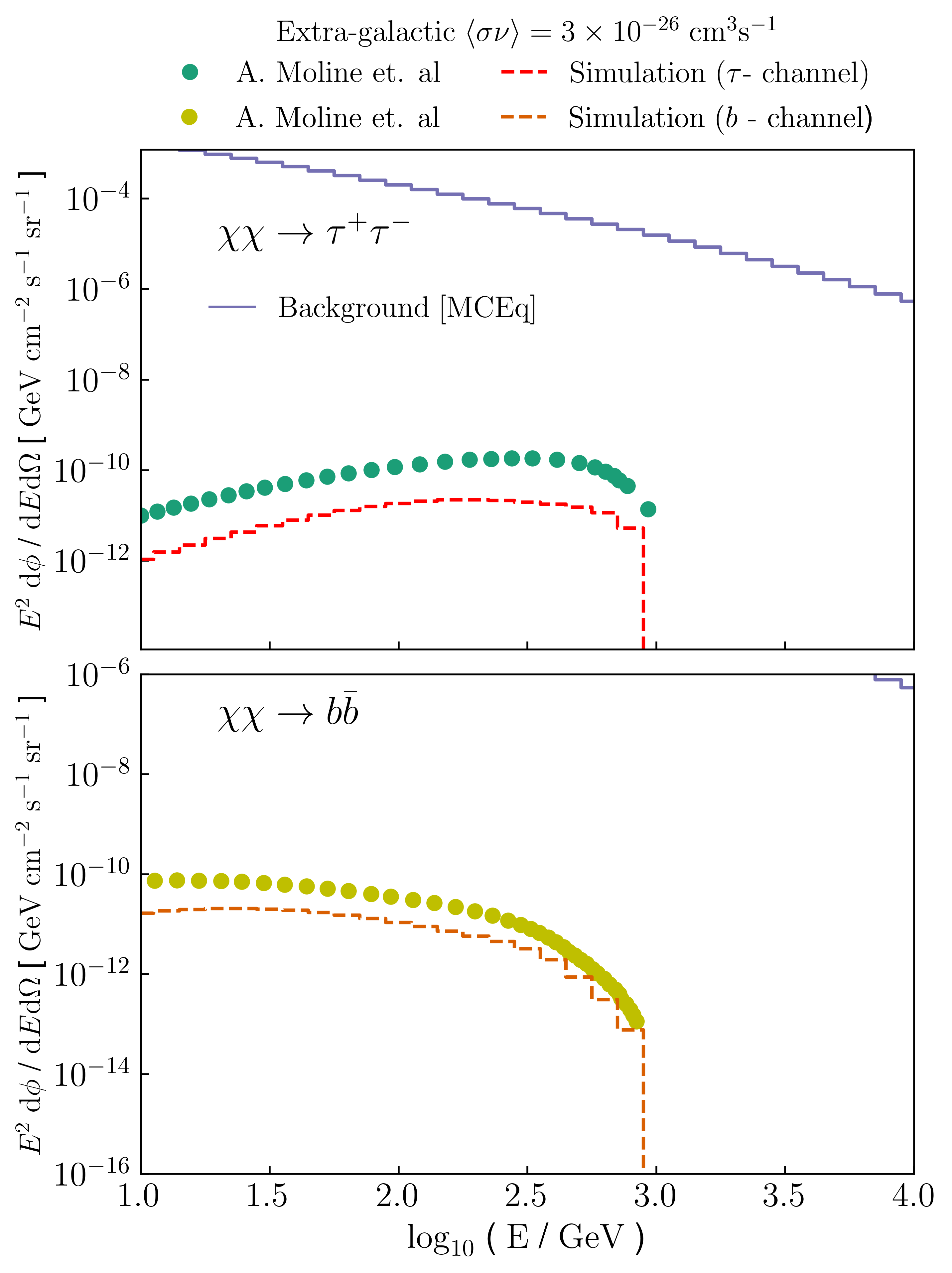}
        \caption{Differential neutrino fluxes of extra-galactic sources via in-direct annihilation of 10TeV-DM pairs into a $\tau^{+}\tau^{-}$ pair with $\left \langle\sigma\nu\right \rangle=3\times 10^{-26}$ $\text{cm}^{3}\text{s}^{-1}$.  The difference in the flux results (dashed) compared to \cite{Molin__2016} (dotted) is due to uncertainties associated with the approximation methods.
        The blue line is MCEq simulated atmospheric neutrino background. Here we have parameterized the red-shift grid with the energy of neutrino at production $E_{\text{prod}}$ and the mass of DM $m_\chi$.}
        \label{fig:flux_extra}
\end{figure}

There are several approximation models for \autoref{eq:G}. Here, we use the one described in \cite{Lopez_}. We give a brief discussion of the different approximations in \autoref{app:extra_}  and \autoref{app:halo} along with a few intermediate results of our calculations.  In \autoref{fig:flux_extra}, we show the flux results for two different DM indirect annihilation channels to neutrinos, $\tau$-lepton(upper) and b-quark(bottom). Here we have used  $\left \langle\sigma\nu\right \rangle$ = $3\times10^{-26}$ $\text{cm}^{3}\text{s}^{-1}$ for comparing the flux results with Fig.5 in reference \cite{Molin__2016}. There is a factor of 10 difference between our results and the \cite{Molin__2016} flux, which can be explained due to differences in the approximation methods. Additionally, \cite{Molin__2016} mentioned they were using redshifts up to z = 49 in their simulations, which differ from our calculations. Even using the higher fluxes from \cite{Molin__2016}, does not change the final CL limits significantly. For the direct annihilation channel to neutrinos, \cite{carlos} showed that the extra-galactic flux would, at maximum contribute 10\% to the final CL limit. We consider the 10\% as an upper bound on contributions by extra-galactic sources. Although the possible extra-galactic contribution could vary between 0.0001\% to 10\% of the galactic one, it would at best minimally improve the final constraint value on the cross-section. Therefore, we neglect the extra-galactic component in the following analysis.

More analyses were done on extra-galactic contribution from the halo boost factor and halo substructures in \cite{Okoli_DM_halo_EXG} and \cite{Ando}. According to \cite{Okoli_DM_halo_EXG}, the total sub-halo contributions are dominant compared to those of smooth halo structures in the high mass regime, approximately above $10^4$ GeV masses. In such scenarios, the contributions from extra-galactic sources boosted by substructures could be comparable with the galactic contribution. Future directional searches targeting nearby galaxies would be ideal for probing such substructure contributions further.

\section{Effective Areas and Background}\label{sec:Background}
\begin{figure}[h]
        \centering
        \includegraphics[width=8.6cm]{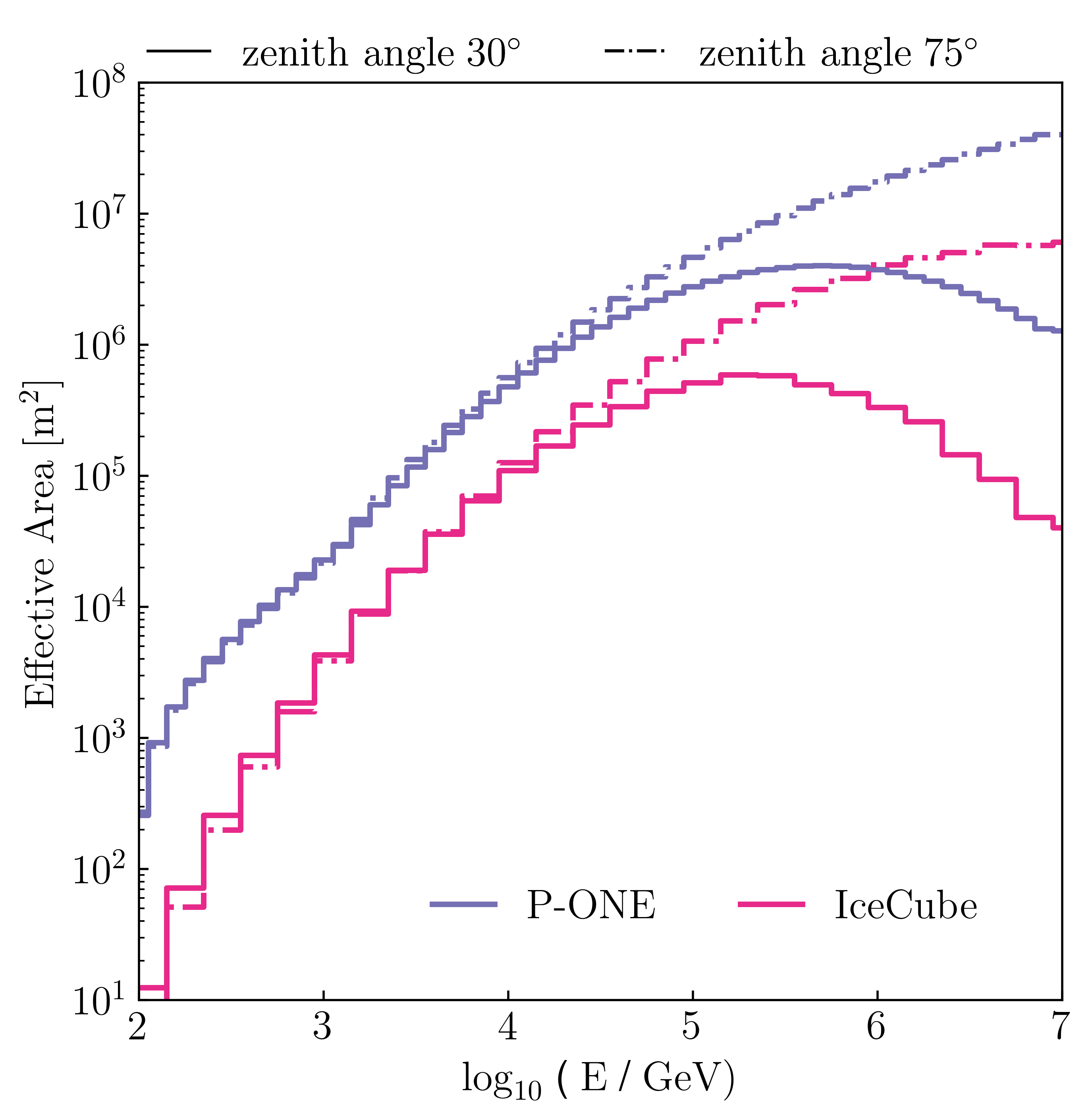}
        \caption{Effective area comparison between IceCube\cite{Ice_data} and P-ONE\cite{PONE} for zenith angles $30^\circ$ and $75^\circ$. Due to the larger geometric volume, P-ONE will possess a larger effective area than IceCube. The solid line style corresponds to $-1<\cos{\theta}<-0.5$, and the dashed-dotted to $-0.5<\cos{\theta}<0.5$. The effective area line for $0.5<\cos{\theta}<1$ would overlap with the $-0.5<\cos{\theta}<0.5$ line.}
        \label{fig:Eff_area}
    \end{figure}

This section is dedicated to the effective areas of the IceCube and P-ONE observatories and the corresponding background fluxes. We describe the flux-to-count conversion with the help of effective areas since the later statistical analysis requires an event rate prediction.
\\
The published IceCube effective areas \cite{Ice_data} are binned with a zenith angle grid between $0^{\circ}$ and $180^{\circ}$. In comparison, the simulated effective areas for P-ONE are differentiated into trimesters in the sky between $0^{\circ}$ to $180^{\circ}$( $-1<\cos{\theta}<1$). In \autoref{fig:Eff_area}, we show the effective areas for both P-ONE and IceCube. The solid line style corresponds to $-1<\cos{\theta}<-0.5$, and the dashed-dotted to $-0.5<\cos{\theta}<0.5$. Note that the effective area line for $0.5<\cos{\theta}<1$ would overlap with the $-0.5<\cos{\theta}<0.5$ line. In addition to the effective area, in IceCube's case, we use the provided mixing matrices, to move from true neutrino energies to smeared energies. For P-ONE, we use the IceCube public energy smearing to approximate the energy reconstruction. The model is described at the end of this section.
Observed and projected background counts are used for the likelihood analysis of the signal. To model the atmospheric background for P-ONE, we use MCEq\cite{MCEq}, with the primary model H4a\cite{H4a} and interaction model SYBILL2.3c \cite{Sybill}. Since the published effective areas are designed for track-like events, we will exclusively use $\nu_{\mu}$ charged-current events.
\\
Along with the atmospheric background, we have to include the astrophysical background for both detectors. We assume a power-law spectrum can model the astrophysical flux
    \begin{equation}
        \frac{d \Phi_{Astro}}{d E} =  \phi_{0}\times \Big(\frac{E}{100\text{ TeV} }\Big)^{-2.53\pm0.07},
        \label{eq:Astor}
    \end{equation}
where we set the parameters to the best-fit values measured by IceCube, with a spectral index of 2.53$\pm0.07$ and $\phi_{0}= 1.66^{+0.25}_{-0.27}\times10^{-18}$  GeV $\text{cm}^2$ s sr$^{-1}$\cite{Aartsen_astro}.
\\
The differential fluxes for the background along with the signal fluxes for the galactic (\autoref{eq:galactic}) signals are then convolved with the effective areas of the neutrino detectors to produce the signal and background counts represented as $N_{events}$ with $t_{run}$ (run time of the detector) in \autoref{eq:counts} 
\begin{equation}\label{eq:counts}
N_{events} =  \iint \frac{d\Phi}{dE} \times A_{\text{eff}}(E, \theta, \phi) d\Omega dE \times t_{run}.
\end{equation}

\begin{figure}
        \centering
        \includegraphics[width=8.6cm]{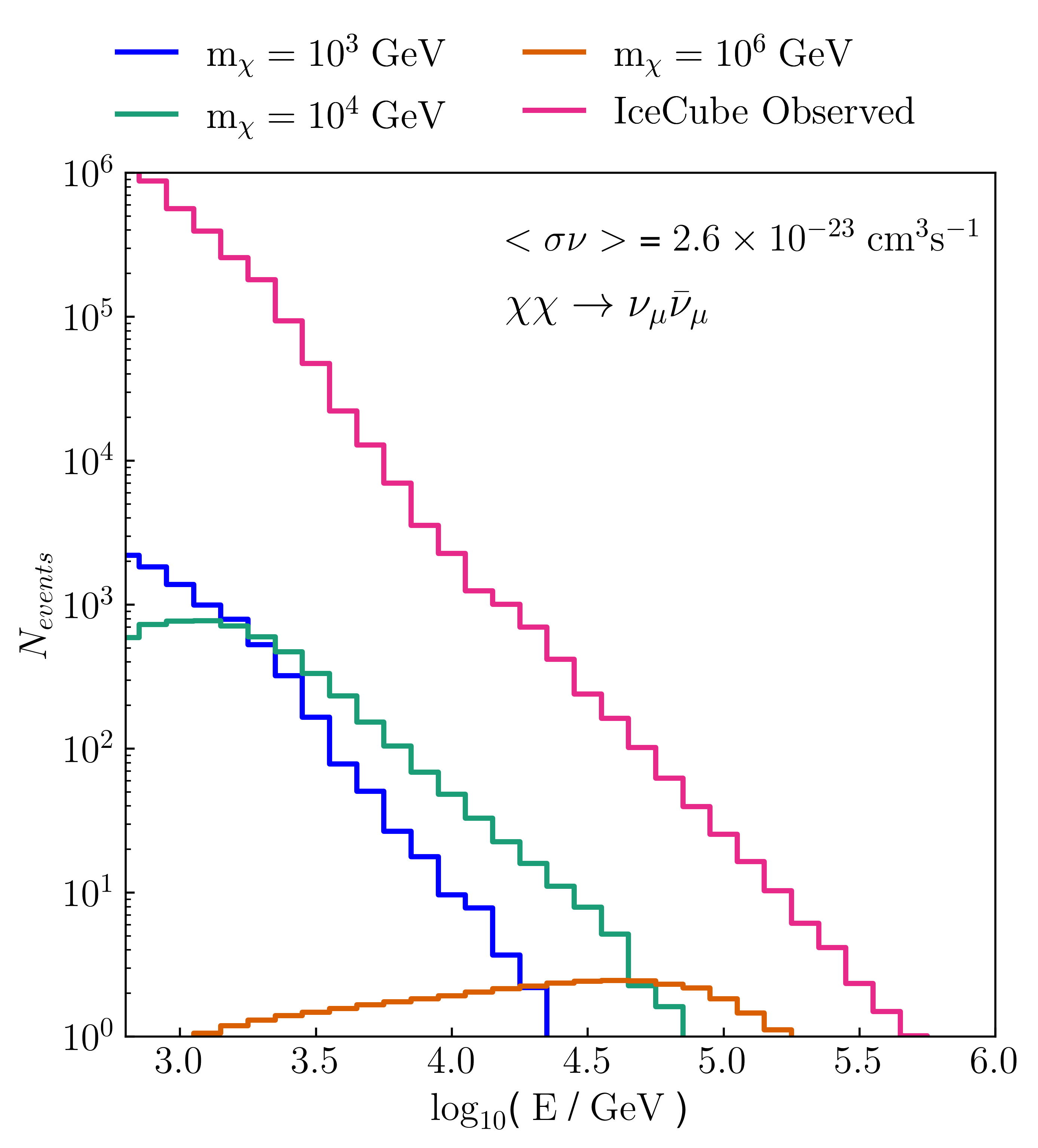}
        \caption{IceCube observed events (pink) and predicted signal counts with dark matter masses of 1 TeV (blue), 10 TeV (green), and 1 PeV (yellow). Here, the thermally averaged cross-section is $\left \langle\sigma\nu\right \rangle=1\times 10^{-23}$ $\text{cm}^{3}\text{s}^{-1}$.}
        \label{fig:counts_ice}
\end{figure}
Using this energy-smearing procedure, we obtain the expected signal counts, depending on the dark matter mass and thermally averaged cross-section. Details on the energy smearing are given in \autoref{app:distro}. In \autoref{fig:counts_ice} we compare ten years of observed IceCube events\cite{Ice_data} with our predicted dark matter signals.
\begin{figure}
    \centering
     \includegraphics[width=8.6cm]{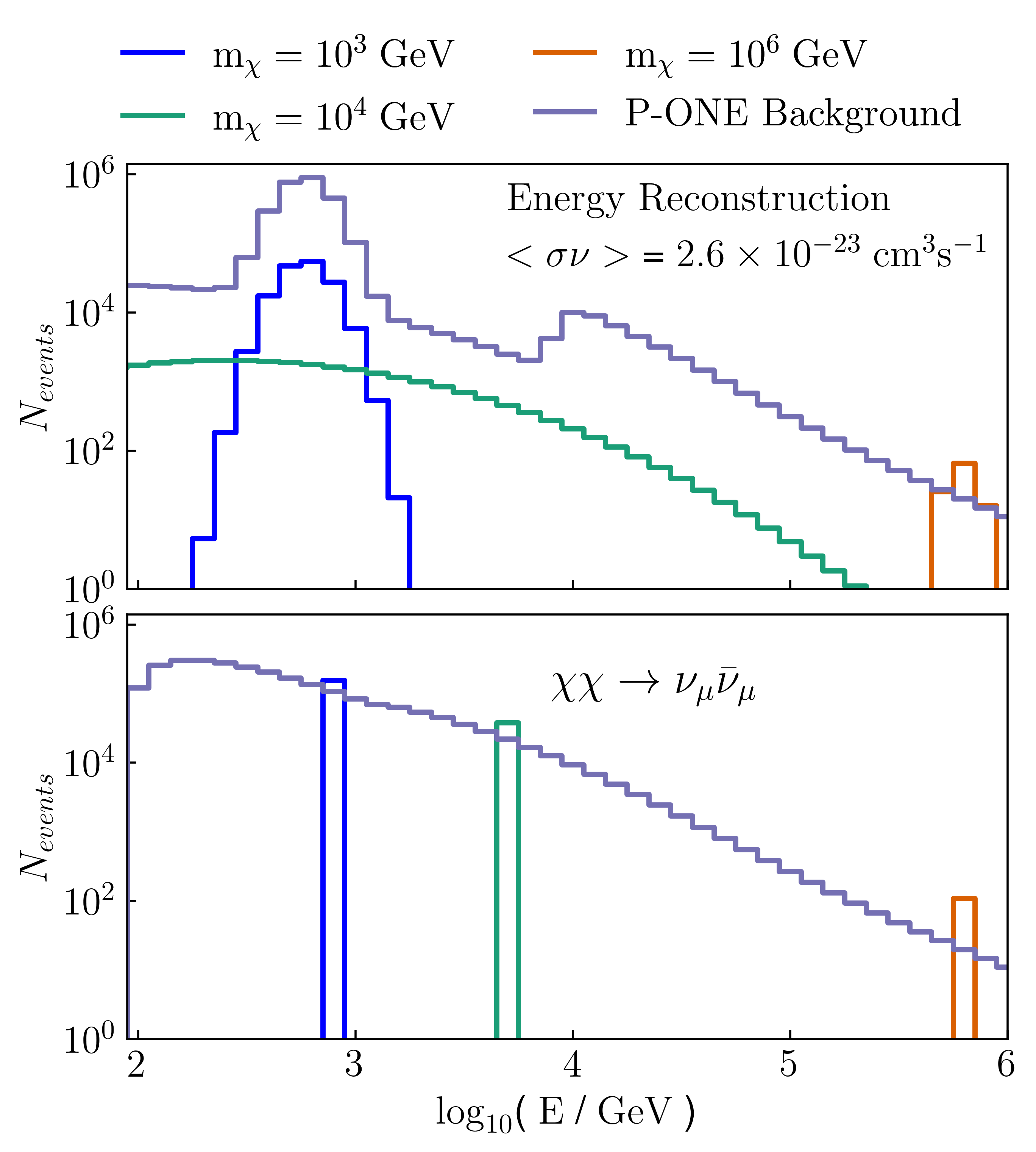}
     \caption{Expected counts for the P-ONE detector with ten years run-time for the direct DM to neutrino annihilation channel. The upper graph depicts expected counts with energy smearing. The bottom graph represents an ideal scenario with perfect energy reconstruction at the detector. The smearing parameters applied are similar to the IceCube public data energy smearing matrix from \cite{ice_recon}. The $10^4$ GeV peak seen in the fluxes is due to the parametrization we chose here.}
    \label{fig:counts_pone}
\end{figure}
Similarly, we show the expected background and dark matter to neutrino signal events after ten years for P-ONE in \autoref{fig:counts_pone}. There we give the results for an IceCube-like energy smearing (top) and a perfect detector (bottom). The energy smearing procedure is done with a log-normal distribution of events. This procedure is briefly explained in \autoref{app:distro}. The parameters used for the smearing are similar to the IceCube parameterization depicted in \cite{ice_recon}. In \autoref{fig:counts_pone}, the smeared flux curves (upper graph) show changes in the energy distribution shape for different mass regimes. This is due to the different parameters according to their true energy region (resonant peaks with the DM mass)  as discussed in \autoref{app:distro}.
\\
\section{Analysis and Results}\label{sec:Limits}
In this section, we perform a log-likelihood ratio test on IceCube data and predicted events for P-ONE. Both of these analyses were done with the NFW profile and combining the signal fluxes as discussed in \autoref{sec:modeling} with galactic contributions. We define the binned likelihood function for the signal hypothesis, $\mathcal{H}_1$, to be
 \begin{equation}
     \mathcal{L}(\mathcal{H}_1) \& = \prod\limits_{i}\mathcal{P}\left(d_i|\mu_{s,i} (\bm{\theta}) + \mu_{atmos, i}(\eta_1) + \mu_{astro, i}(\eta_2)\right)
 \end{equation}
The binning used for the background and signal distributions are the same. Here, $\mathcal{P}$ is a Poisson distribution, $i$ runs over the energy bins, $d_i$ is the measured event count taken from data and $\mu_{s, i}$ is the expected event count given the DM parameters $\bm{\theta}$. $\mu_{atmos, i}$ and $\mu_{astro, i}$ are the expected atmospheric and astrophysical events depending on nuisance parameters $\eta_1$ and $\eta_2$. For the astrophysical flux, $\eta_2$ is set to reflect the uncertainties on the parameters given in \autoref{eq:Astor}, while for the atmospheric background, we assume a flat 20\% uncertainty. A comparison between the CR, astrophysical flux uncertainties, and $\eta_i$, is shown in \autoref{app:background_uncertainties}. We define the null hypothesis, $\mathcal{H}_0$, to have zero signal events ($\mu_{i,s}(\bm{\theta})=0$). We then define the test statistic $q_\mu$ as the log-likelihood ratio
 \begin{equation}
     q_\mu = -2\log\frac{\mathcal{L}(\mathcal{H}_1)}{\mathcal{L}(\mathcal{H}_0)}.
 \end{equation}
Since this is a diffuse analysis, we can disregard the right ascension of the events and create a background model by scrambling data in the right ascension. Note, that this reduces the sensitivity of this analysis since we remove any possible benefits from potential DM overabundance in the galactic center. To create the background model, we scrambled 10 years of IceCube data $10^6$ times and constructed the mean binned in energy. Then we fit our atmospheric and astrophysical neutrino fluxes to the mean by scaling our calculations, see \autoref{sec:Background}. The scaling values are 1.1 for the atmospheric flux and 0.98 for the astrophysical flux. We then add the previously mentioned nuisance parameters $\eta_1$ and $\eta_2$ as floating normalizations to these best fits. We then follow \cite{doi:10.1146/annurev.nucl.57.090506.123052} and marginalize over these parameters, to minimize the resulting limits. This is to account for the signal possibly contributing to the background.
 
\begin{figure}
    \centering
     \includegraphics[width=8.6cm]{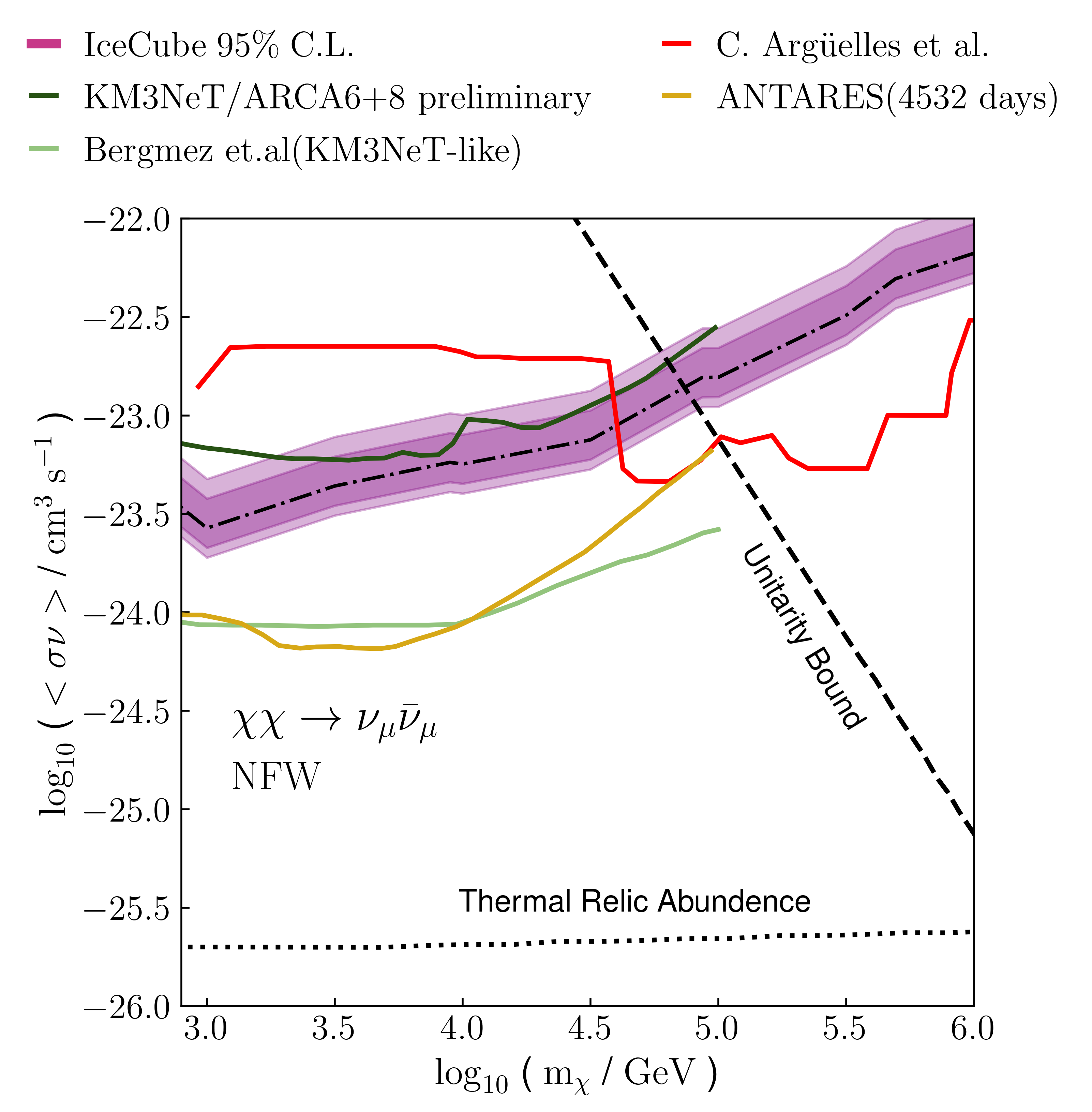}
     \caption{IceCube 95\% C.L. limits using ten years of public data on the $\nu_\mu \bar{\nu}_\mu$ annihilation channel (black). The dark pink and light pink bands represent uncertainties due to the dark matter density and further model-related uncertainties. The solid red curve represents IceCube's high energy sensitivity estimate from \cite{carlos} (red). The solid dark green and yellow curves are the current bounds set by KM3NeT \cite{km3net_arca6} and ANTARES \cite{ANTARES_4532}. We also show light green curve for Bergmez et.al. \cite{KM3NET_bergmes}(KM3NeT-like), thermal relic abundance \cite{thermal_relic_var} (black dots), and the unitarity bound (black dashed).}
       \label{fig:limits_ice}
\end{figure}

In \autoref{fig:limits_ice}, we show the resulting confidence level (C.L.) limits for direct annihilation of DM pairs to $\nu_{\mu}\bar{\nu}_{\mu}$ using ten years of public IceCube data. The red solid curve represents the 95\% C.L. sensitivity predicted in \cite{carlos} for the high $m_{\chi}$ region with a Background Agnostic method. The green line shows the current bound set by KM3NeT \cite{km3net_arca6} using a fraction of the planned detector. In yellow, we show the bounds set by ANTARES \cite{ANTARES_4532}. In \autoref{app:analysis_test}, we compare these results to the expected ones, as well as a comparison of best-fit and injected background and signal events.

The strong bound set by ANTARES is due to its location and the analysis method. While far smaller than IceCube, the galactic center lies in ANTARES' most sensitive region. This allows ANTARES to perform a directional study of the galactic center, boosting its sensitivity. \cite{ice_gc_penalty} shows that the location leads to approximately a factor of 20  sensitivity penalty for IceCube when performing similar analyses.

In \cite{KM3NET_bergmes} the potential of a full KM3NeT-like \cite{km3net_eff_bergmes} detector is shown calculated with an Angular Power Spectrum (APS) method assuming a run time of ten years. This would push the KM3NeT line below the ANTARES line.

\begin{figure}
    \centering
    \includegraphics[width=8.6cm]{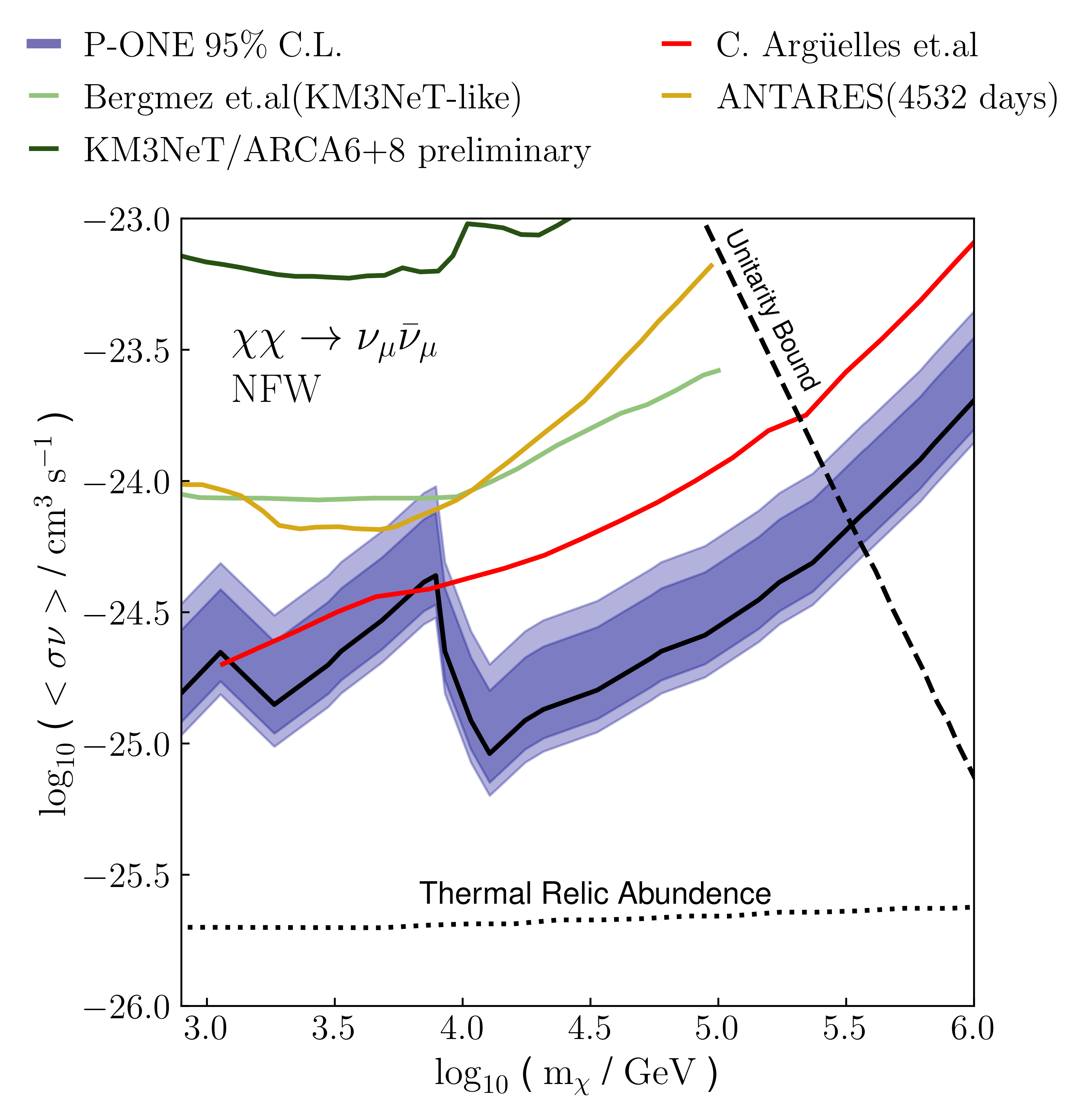}
    \caption{Sensitivity estimation for P-ONE. The solid black line shows the $95\%$ C.L. limit. The 90\% (dark blue) and 68\% (light blue) bands represent uncertainties due to the dark matter density. The sensitivity is estimated with ten years of run-time, including energy smearing. The solid red curve represents the previous results from \cite{carlos}, the green curve represents the results from KM3NeT\cite{km3net_arca6}, and the yellow those from ANTARES \cite{ANTARES_4532}. We also show light green curve for Bergmez et.al. \cite{KM3NET_bergmes}(KM3NeT-like).  The constraint linear shows a jump in the $10^3$ GeV to $10^{3.7}$ GeV energy region. This is due to the log-normal distribution used in our P-ONE energy-smearing model (see \autoref{app:distro}).} 
    \label{fig:limits_pone}
\end{figure}

We show the resulting predicted sensitivities for P-ONE in \autoref{fig:limits_pone}. The stark change in shape between 1 and 10 TeV mass is due to the energy-smearing model applied here. This change is especially relevant in the direct annihilation channel to neutrinos due to the expected peak at the mass resonance. If one removes our energy reconstruction assumptions, the P-ONE sensitivities will improve greatly, following a similar shape as the red line.
\\
The P-ONE sensitivity could be improved by performing a spatial analysis similar to \cite{KM3NET_bergmes}. Purely from a scaling perspective, when comparing the effective areas, we would expect the P-ONE sensitivity to improve further by a factor between five and ten when utilizing spatial information.

In \autoref{app:conf_ex}, we show example test statistic distributions and the resulting confidence limits for P-ONE and IceCube for the scanned model parameters, $m_\chi$ and $\left \langle\sigma\nu\right \rangle$.

\begin{figure}
       \centering
       \includegraphics[width=8.6cm]{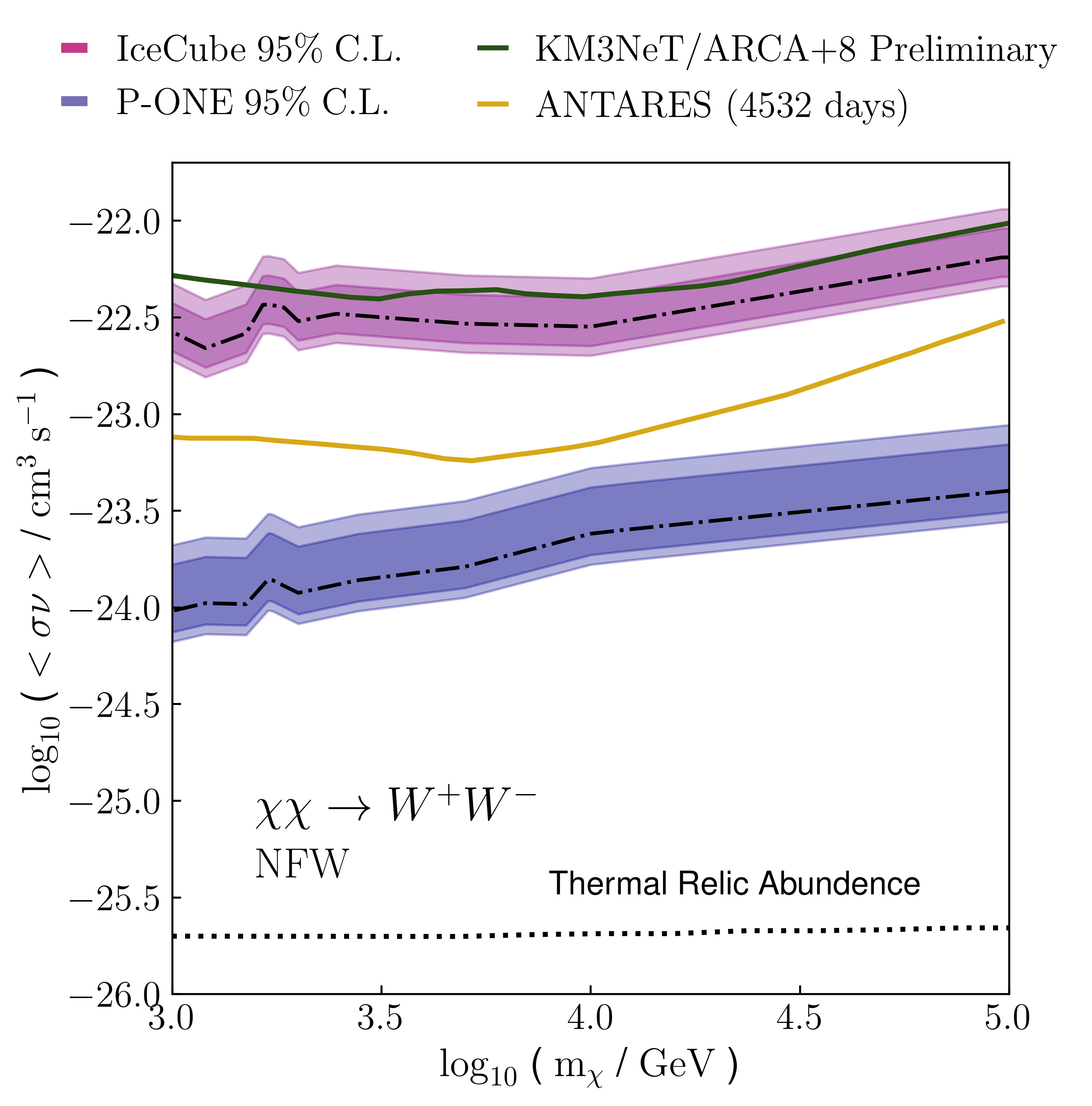}
       \caption{Thermally-averaged WIMP annihilation cross section as a function of the WIMP mass for $W^{+}W{-}$ annihilation channels. The limits for IceCube (pink band) and estimated sensitivity P-ONE (purple band) are shown. Two $W^{+}W{-}$ results from ANTARES(4532 days)\cite{ANTARES_4532} with yellow curve and KM3NeT/ARCA6 \cite{km3net_arca6} are also presented as a comparison with dark green. }
       \label{fig:limits_w}
\end{figure}

We now analyze DM pair annihilation to W-bosons and consequent decay to muon neutrinos, depicted in \autoref{fig:limits_w}. This analysis is done for the P-ONE detector with the numerical data from \cite{Cook_book} assuming equilateral distribution amongst the flavors. The low energy cutoff was lowered to 500 GeV for this analysis. Our new limits (in the case of IceCube) are in energies above 1 TeV more stringent than previous analyses, e.g., IC86 by IceCube\cite{IC86_Aartsen_2016}. At the same time, we predict P-ONE's sensitivity using ten years of data to be even greater.

\begin{figure}
    \centering
    \includegraphics[width=8.6cm]{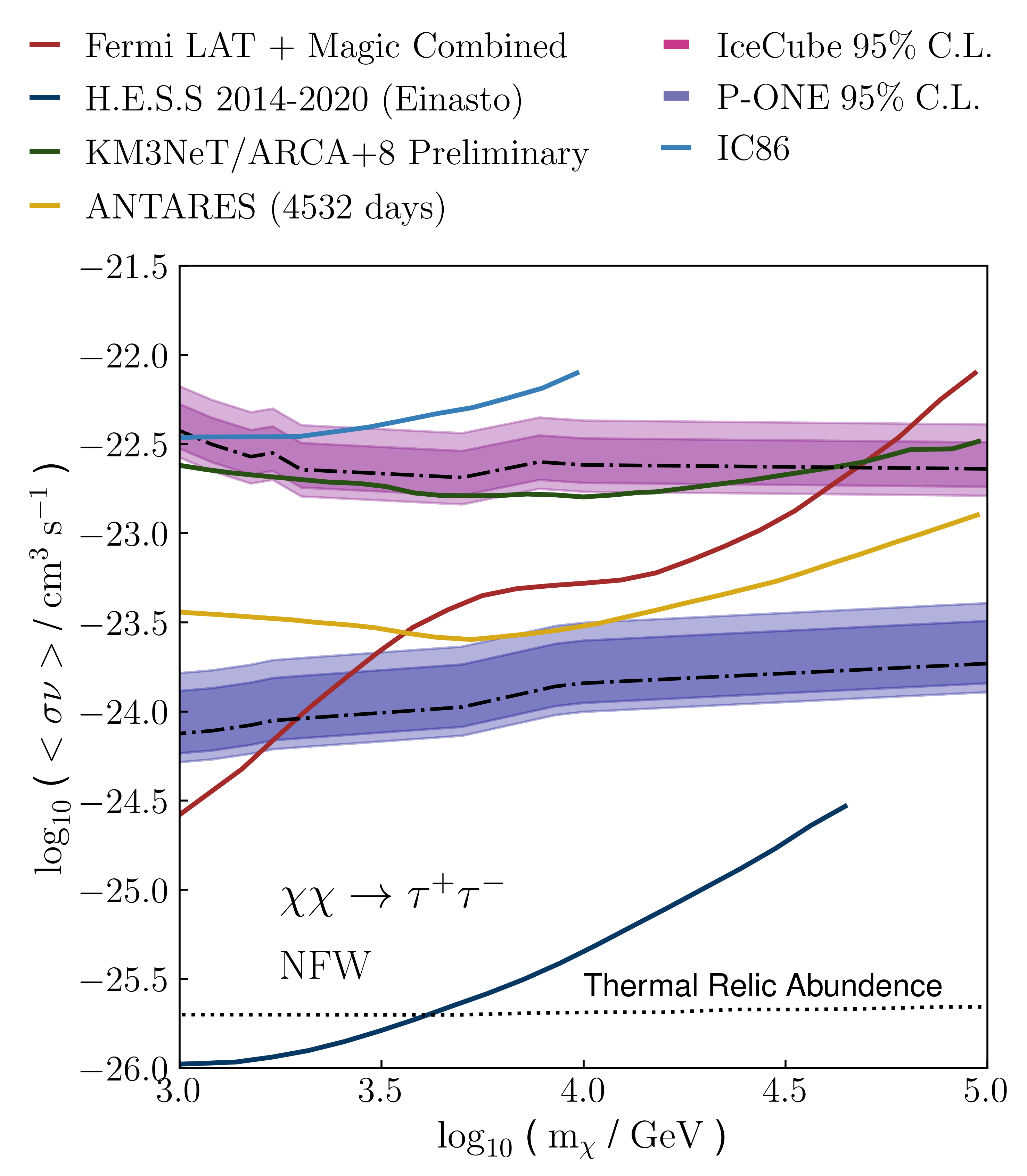}
    \caption{Thermally-averaged WIMP annihilation cross section as a function of the WIMP mass for $\tau^{+}\tau{-}$ annihilation channels.  The limits for IceCube (pink band) and estimated sensitivity P-ONE (purple band) are shown. Several $\tau^{+}\tau{-}$ results from other neutrino experiments are shown: ANTARES\cite{ANTARES_4532}(yellow) galactic plane search, IceCube 86 strings Galactic halo all-sky cascades\cite{IC86_Aartsen_2016} (blue) and KM3NeT/ARCA6 + 8 preliminary galactic center search\cite{km3net_arca6} with dark green curve. The latest upper limits from gamma-ray combined searches by Fermi-LAT and MAGIC (brownish-red) on 15 dwarf satellite galaxies \cite{Fermi_magic_2016}, H.E.S.S galactic center search \cite{GC_Hesse} with dark blue curve.}
    \label{fig:limits_pone_tt}
\end{figure}

Similarly, we have analyzed the $\tau$-lepton channel for IceCube and P-ONE, with an NFW profile for $\nu_{\mu}\bar{\nu}_{\mu}$ pair production. \autoref{fig:limits_pone_tt} shows the improvement between our results and the previous  IC86 Galactic halo with all sky cascade signal\cite{IC86_Aartsen_2016} study.
\\
In both annihilation scenarios, we expected P-ONE to outperform IceCube by an order of magnitude. The "jump" in the $10^3$ to $10^{3.5}$ energy region which one can see in \autoref{fig:limits_pone}, is still present in \autoref{fig:limits_w} and \autoref{fig:limits_pone_tt}. However, the jump is not as intense as it is for the direct channel since the neutrino spectrum is not a mass resonant peak rather, it is a distribution over the energy up to the dark matter mass discussed in \autoref{fig:spectrum_1e3}. 

\autoref{fig:limits_pone_tt} shows that the projected P-ONE sensitivity to dark matter annihilation will exceed the constraints set by Fermi+MAGIC $\gamma$-ray limits for 15 dwarf satellite galaxies \cite{Fermi_magic_2016} in the high-mass region as well as exceeds that of predicted by ANTARES collaboration for 4532 days \cite{ANTARES_4532}. 

\section{Conclusion}\label{sec:conculsion}
We performed dark matter annihilation searches on ten years of public IceCube data, setting the most stringent constraints on DM self-annihilation to neutrinos in the high-mass regime. Compared to previous analyses, the constraints set here also show almost one order of magnitude improvement to previous neutrino studies for the galactic halo, galactic center, and extra-galactic diffused sources in both the direct and indirect annihilation channels.
\\
We also modeled sensitivities for a new proposed neutrino telescope, P-ONE. These show even greater potential than the constraints derived in this work for IceCube and can compete with constraints set by Fermi-LAT gamma-ray experiments. This indicates that P-ONE will play an important role in future DM searches, especially in the 10-100 TeV range.
\\
We expect its detection potential could be pushed further toward the thermal relic abundance when performing an analysis similar to \cite{KM3NET_bergmes}, which would include spatial information, unlike the diffuse analysis presented here. Directional information may become especially relevant when analyzing individual galaxies and studying sub-halo contributions\cite{andosubhalo} to the neutrino flux.

\section{Acknowledgments}
We thank Dr. Christian Haack and Dr. Martin Wolf for their valuable discussions, as well as Prof. Dr. Alejandro Ibarra and Dr. Shin'ichiro Ando for their guidance. Especially, We would also like to thank the support from Prof. Dr. Elisa Resconi. This research was supported by the German National Science Foundation (Deutsche Forschungsgemeinschaft, DFG) under the umbrella of the Special Research Activity (Sonderforschungsbereich, SFB1258) and the Australian Research Council's Discovery Projects funding scheme (Project DP220101727).
\nocite{*}

\bibliographystyle{apsrev}
\bibliography{bibliography}

\appendix
\renewcommand\thefigure{\thesection.\arabic{figure}}
\section{Spectrum for DM Indirect Annihilation Channels}\label{app:spectrum}

The spectrum data we used for indirect DM annihilation channels (at sources) are available at \cite{Cook_book} and we show them in \autoref{fig:spectrum_1e3}
The produced neutrinos are expected to have a long enough propagation distance so that after neutrino oscillation, they have a 1:1:1 flavor ratio at the Earth.
\setcounter{figure}{0}
\begin{figure}[H]
\centering
\includegraphics[width=8.6cm]{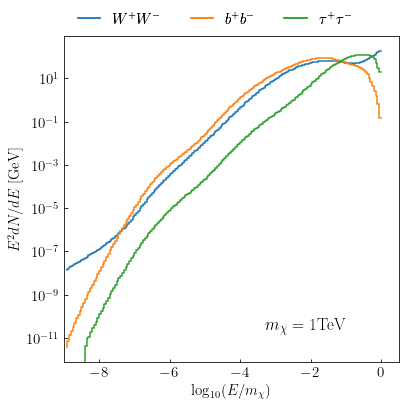}
\caption{Energy spectrum $E^2 dN/dE$ respect to $x=E/m_{\chi}$) with $m_{\chi} = 10^{3}$ GeV. Shown are the W-boson (blue), b-quark (orange), and $\tau$-lepton (green) channels.}
\label{fig:spectrum_1e3}
\end{figure}

\section{Extra-Galactic Contributions}\label{app:extra_}
The extra-galactic flux is particularly interesting and rarely studied since the neutrinos produced in the extra-galactic sources would suffer non-negligible redshift effects. The neutrino flux from extra-galactic sources is calculated by Eq. \autoref{eq:Extra} for DM direct annihilation to the neutrino pair channel
\begin{equation}\label{eq:Extra}
    \begin{split}
    \frac{d\Phi_{extra}}{d E_{\nu}} &= \frac{1}{4\pi} \frac{\Omega_{DM}^2 \rho_{c}^2 \left \langle\sigma\nu\right \rangle}{\kappa m_{\chi}^2}\frac{1}{3} \\
    & \times\int_{0}^{z_{up}} dz\frac{[1+G(z)](1+z)^3}{H(z)}\frac{dN_{\nu}}{dE_{\nu}}.
    \end{split}
\end{equation}
Where $H(z)=H_{0}[ (1+ z)^3 \Omega_{m} + (1 + z)^4 \Omega_{r} + \Omega_{\Lambda}]^{1/2}$ is the
time-dependent Hubble parameter, $\rho_{c}$ is the critical density of
the Universe, and $\Omega_{m}$, $\Omega_{r}$, and $\Omega_{\Lambda}$ are, respectively, the
fractions of $\rho_{c}$ made up of matter, radiation, and dark energy\cite{carroll2004spacetime}. The number spectrum of neutrino pair production is given in \autoref{eq:dN_extra}
\begin{equation}\label{eq:dN_extra}
   \begin{split}
    \frac{dN_{\nu}}{dE_{\nu}}&=2\frac{m_{\chi}}{E'^2}\delta \left( \frac{m_{\chi}}{E'}-1 \right) \\
    &=\frac{2}{E}\delta \left[ z-\big(\frac{m_{\chi}}{E} - 1 \big)\right],
    \end{split}
\end{equation}
 which is similar to \autoref{eq:dn_gal}. However, the redshift needs to be considered, which results in an energy transformation. The energy parameterization at Earth is $E = E'(1+z)$, with the energy at the source $E'$ and the redshift, $z$, of the extra-galactic source. Similarly, the numerical spectra from \cite{Cook_book} also require the transformation factor of $1/(1+z)$ for each energy bin.

$G(z)$ is the halo boost parameter at redshift $z$ describing the clustering effect of matter in a galaxy's halo, given by
\begin{equation}\label{eq:G}
    \begin{split}
    G(z) &= \frac{1}{\Omega_{DM,0}^2 \rho_{c}^2(1+z)^6}\\
    & \times \int dM \frac{dn(M,z)}{dM}\int dr 4\pi r^2 \rho_{\chi}^2(r).
    \end{split}
\end{equation}
With the halo boost, the DM annihilation rate has been parameterized with respect to redshift $z$ and halo mass $M$. $dn/dM$ describes the number distribution of halo masses and is strongly related to the halo mass function (HMF)\cite{watson}. A detailed discussion about its calculation is provided in \autoref{app:halo}. The selection of minimum halo mass affects the uncertainties of the final integral result. The smaller halos are more concentrated and contribute more to the neutrino flux, so choosing the lower integration bound is important. In this calculation, we set $M_{min} = 10^{-3} M_\odot\ $ as a conservative lower limit\cite{M_halo_cornell}, \cite{M_halo_shoemaker}.
\\
\section{Halo Annihilation Boost-factor}\label{app:halo}
\begin{figure}
\centering
\includegraphics[width=8.6cm]{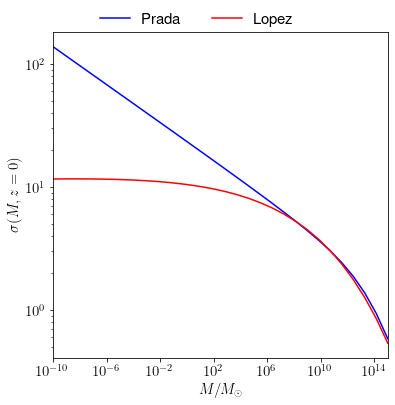}
\caption{Dependence of the linear density field $\sigma$, on the Mass, M. The results from \cite{Prada_2012} and \cite{Lopez_} have been compared here. The deviation is hard in the low halo mass region, whereas in the higher mass region, both approximations converge.}
\label{fig:sigma_fun}
\end{figure}

The halo boost factor G(z) depends on the halo mass function, which itself has a dependence on the variance of linear field density $\sigma$\cite{halo_func}
\begin{equation}\label{eq:sigma2}
	\sigma^2 = \Big( \frac{D(z)}{D(0)} \Big)^2 \int \frac{dk}{k} \frac{k^3 P(k)}{2\pi^2}|\tilde{W}(kR)|^2.
\end{equation} 
 \autoref{eq:sigma2} defines the $\sigma$. The $z$ dependence of $\sigma$ is from the growth factor D(z)\cite{halo_func}. The $\tilde{W}$(kR) is the top-hat filter function and P(k) is the power spectrum. These two parts have only M dependence but not z. Therefore, we can treat M,z-dependence of $\sigma$ separately. For $\sigma(M, z=0)$ we use the parameterization described in the appendix of \cite{Lopez_} and as in \autoref{eq:sigma}:
\begin{equation}\label{eq:sigma}
	\ln{\sigma^{-1}} = 0.2506 \big( M^{0.07536} - 2.6M^{0.001745} \big).
\end{equation}
The \autoref{eq:sigma} does not include the $z$ dependence of $\sigma$. Its $z$ dependence is introduced with the growth factor $D(z)$. In \autoref{fig:sigma_fun} we compared the approximation from \cite{Lopez_} and from \cite{Prada_2012} at a wide mass range. At the higher mass, the two approximations converge. However, at the low mass region, there seems a significant divergence.

\begin{figure}[H]
\centering\includegraphics[width=8.6cm]{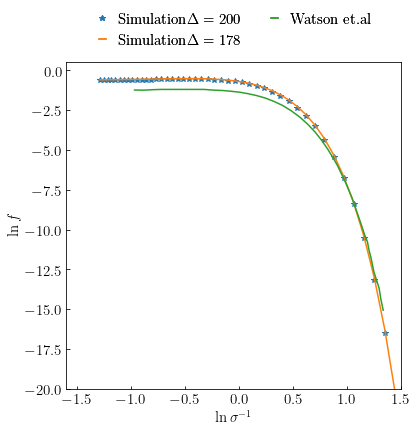}
\caption{The fitted functions for the differential mass function $f$ compared to $\sigma$ at z=0 for both $\Delta$ parameters. The results from Watson et al.\cite{watson} and Lopez et al.\cite{Lopez_} are within a difference of one order of magnitude.}\label{fig:lnf_lnsigma}
\end{figure}

The mass function is non-trivially dependent on $\sigma$ via function f($\sigma$) as shown in \autoref{eq:hmf}\cite{halo_func} 
\begin{equation}\label{eq:hmf}
	\frac{dn}{dM} = f(\sigma)\frac{\rho_{m}}{M}\frac{d\ln(\sigma^{-1})}{dM}.
\end{equation}
$f(\sigma)$ function is called the differential mass function, which has various fitting methods developed over the years, as in \cite{watson}, \cite{Sheth}, and \cite{Reed}, etc. In our analysis, we used the approximation described in \cite{Lopez_} because it is comparable with the results from \cite{watson}, as shown in \autoref{fig:lnf_lnsigma}. The $\Delta$ symbol means the density described in terms of critical density $\rho_{c}$ multiplied with a constant $\Delta$. 

The integral with $\rho_{\chi}$(DM halo mass density) can be approximated by its concentration parameter $c = \frac{r_{\Delta}}{r}$ as in \autoref{eq:rho_inte_x}:
\begin{equation}\label{eq:rho_inte_x}
	\int_{0}^{r_{\Delta}} dr4\pi^2 \rho_{\chi}^2(r) = \frac{M \Delta \rho_{c}(z)}{3}\tilde{g} (c_{\Delta}) .
\end{equation}
The integral has been described as a clumsiness factor in \cite{clumps} or as an enhancement factor described in \cite{ibarra} as well as in \cite{Lopez_}. The enhancement factor is dependent on the concentration parameter, $c_{\Delta}$, and the halo mass, $M$. The $\tilde{g}$ has been approximated as \autoref{eq:tilde_g}\cite{Lopez_}
\begin{equation}\label{eq:tilde_g}
\tilde{g}(c_{\Delta}) = \frac{c_{\Delta}^3}{3} \frac{[1-(1+c_{\Delta})^{-3}]}{[\ln{(1+c_{\Delta})} - c_{\Delta}(1+c_{\Delta})^{-1}]^{2}} \,.
\end{equation}
The concentration parameter can be bounded by the upper limit of 100, hence, one has a limit on the enhancement factor. There are several extensive models to approximate the concentration parameter described in \cite{Prada_2012}, \cite{NFW_better_5}, \cite{Bullock_2001_c}, \cite{Ludlow_2014_c} and \cite{Wechsler_2002_c}. 
\begin{figure}[h!]
\centering\includegraphics[width=8.6cm]{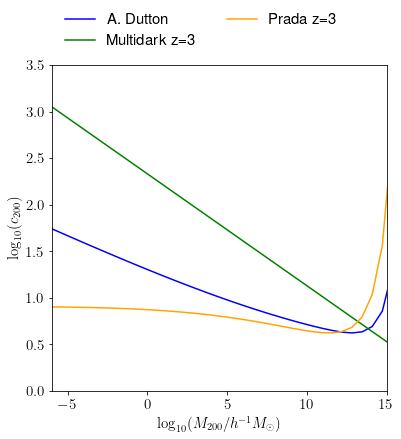}
\caption{The concentration parameter comparison at z=3 throughout the halo mass for various models from \cite{Prada_2012}, \cite{NFW_better_5} and \cite{multidark_c}. The approximations vary in the lower mass region by a greater margin. However, in the high-energy region, the variation is small.}\label{fig:c_2}
\end{figure}

\begin{figure}[h!]
\centering\includegraphics[width=8.6cm]{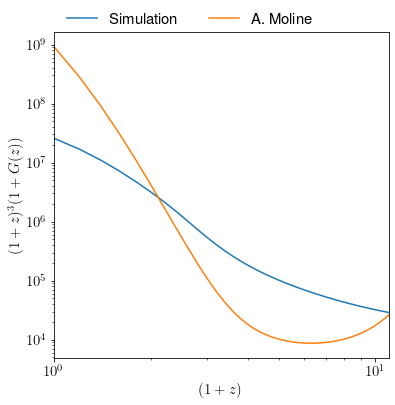}
\caption{The halo boost factor G with respect to redshift z range (0, 20). A comparison between Prada et al.\cite{Prada_2012} simulated result and A. Moline et al.\cite{ibarra}   simulated result is shown here. Both curves deviate to an even larger throughout the larger redshift range. }\label{fig:G_}
\end{figure}

In our analysis, we have opted for the approximation of the concentration parameter $c$ from \cite{Prada_2012}.  The validity ranges over halo mass and redshift for various approximations change drastically, increasing the deviations even more. This results in larger differences in the estimations of halo boost factors and consequently affects the flux estimations. In \autoref{fig:G_} we compared the results of the halo boost factor with the approximation from \cite{ibarra} and simulation from \cite{Prada_2012}. 
\\
The halo boost factor calculated from \autoref{fig:G_} will then be used in \autoref{eq:Extra} to calculate the differential flux. 

\section{Smearing to approximate the Energy Reconstruction}\label{app:distro}
The energy of an incoming particle (true energy) is converted into the optical module-registered photon. The reconstruction procedure converts the registered photon into an energy distribution. The process is specific to each detector. Since P-ONE is still in its early developing stage, we have used a log-normal distribution for the P-ONE energy reconstruction with the parameters close to the IceCube smearing matrix parameters published in \cite{ice_recon}. 
\begin{figure}[H]
    \begin{center}
    \includegraphics[width=8.6cm]{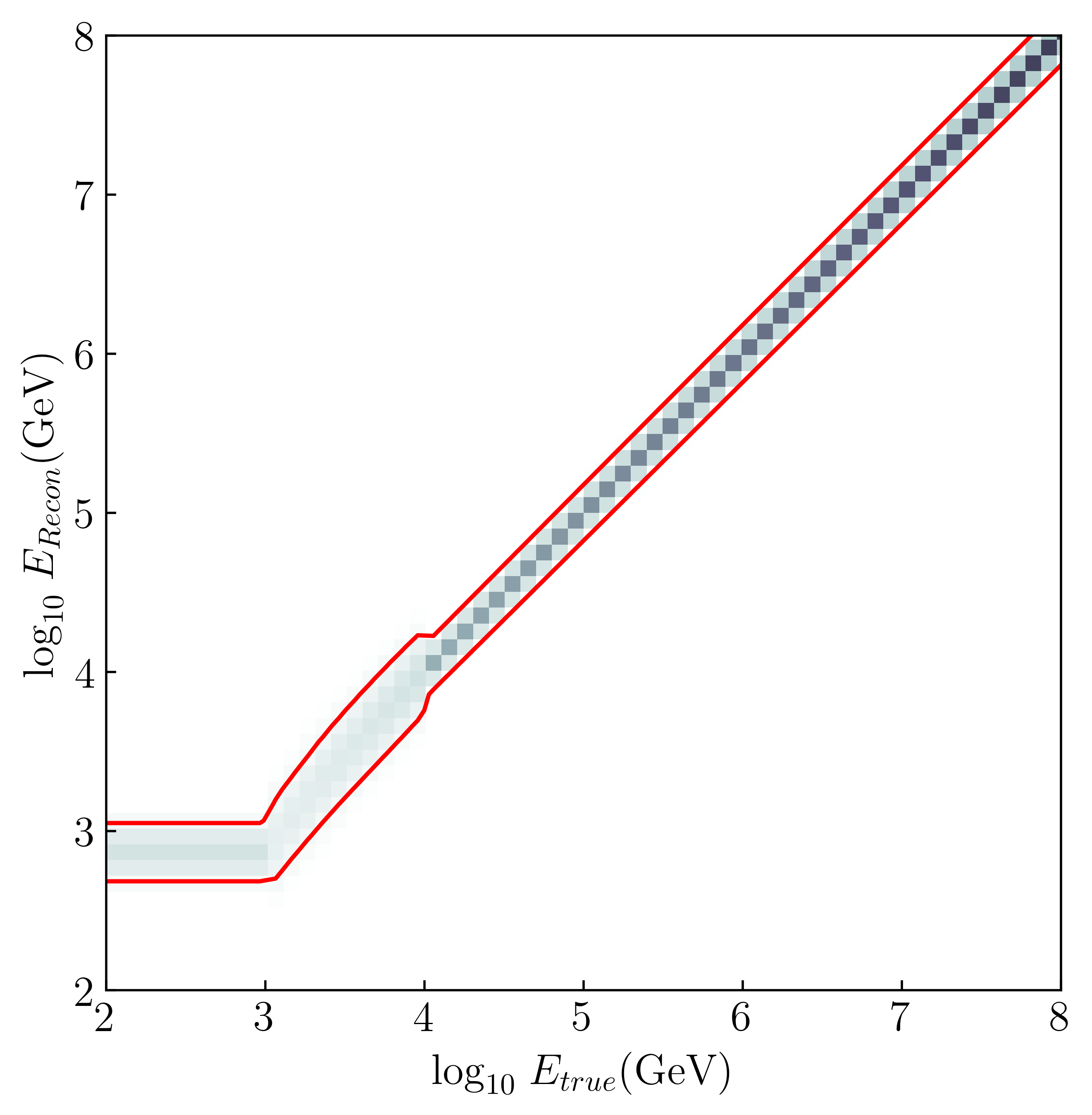}
    \caption{"Energy reconstruction" PDF used for P-ONE. It is a combination of log-normal distributions, with parameters shown in \autoref{tab:E_recon}. The color gradient of gray pixels presents the possibility of 0 (light) to 1 (dark) for a true energy to be reconstructed as the respective smeared energy. The red line indicates the 68\% confidence interval.}
    \label{fig:E_distro}
    \end{center}
\end{figure}
With the help of \autoref{eq:recon} we "reconstruct" the energy distribution of the detected neutrinos. 
\begin{equation}\label{eq:recon}
f(E_{recon}) = \frac{1}{\sigma\sqrt{2\pi}}e^{-\frac{(E_{true}-\mu)^2}{2\sigma^2}} ,
\end{equation}
The PDF describes the smeared energy($E_{recon}$) distribution probability of each True energy ($E_{true}$) bin throughout the energy grid. The \autoref{fig:E_distro} shows the smeared energy distribution ($E_{Recon}$) with respect to the true energy ($E_{true}$) with the parameters described in \autoref{tab:E_recon}. We are using three different “Energy Reconstruction” parameters from low to high energy regions, i.e. below 1 TeV between 1-10 TeV and above 10 TeV. This is modeled in such a way to emulate “IceCube-like” energy smearing behavior as parameterized in  \cite{ice_recon}.  In the \autoref{fig:E_distro}, we see the log-normal distribution's behavior changing drastically from $10^3$ GeV. This change results in the peak observed in the P-ONE limits between the energy band of 1 to 10 TeV. The decline from the peak also occurs due to a change in log-normal distribution at 10 TeV.

\begin{table}
    \centering
    \begin{tabular}{c|c|c|c}
        [TeV] & $E_{true}< 1$ & $1 \leq E_{true} \leq 10$ & $10 < E_{true}$\\
        \hline
        $\mu$ & 0.7 & linear spline & $E_{true}$ \\
        $\sigma$ & 0.45 & linear spline & 0.35 \\      
    \end{tabular}
    \caption{Reconstruction parameters for P-ONE for different true energies given in TeV. The linear spline describes the linear extrapolation between the 1 TeV to 10 TeV boundaries.}
    \label{tab:E_recon}
\end{table}
\begin{figure}[H]
    \begin{center}
    \includegraphics[width=8.6cm]{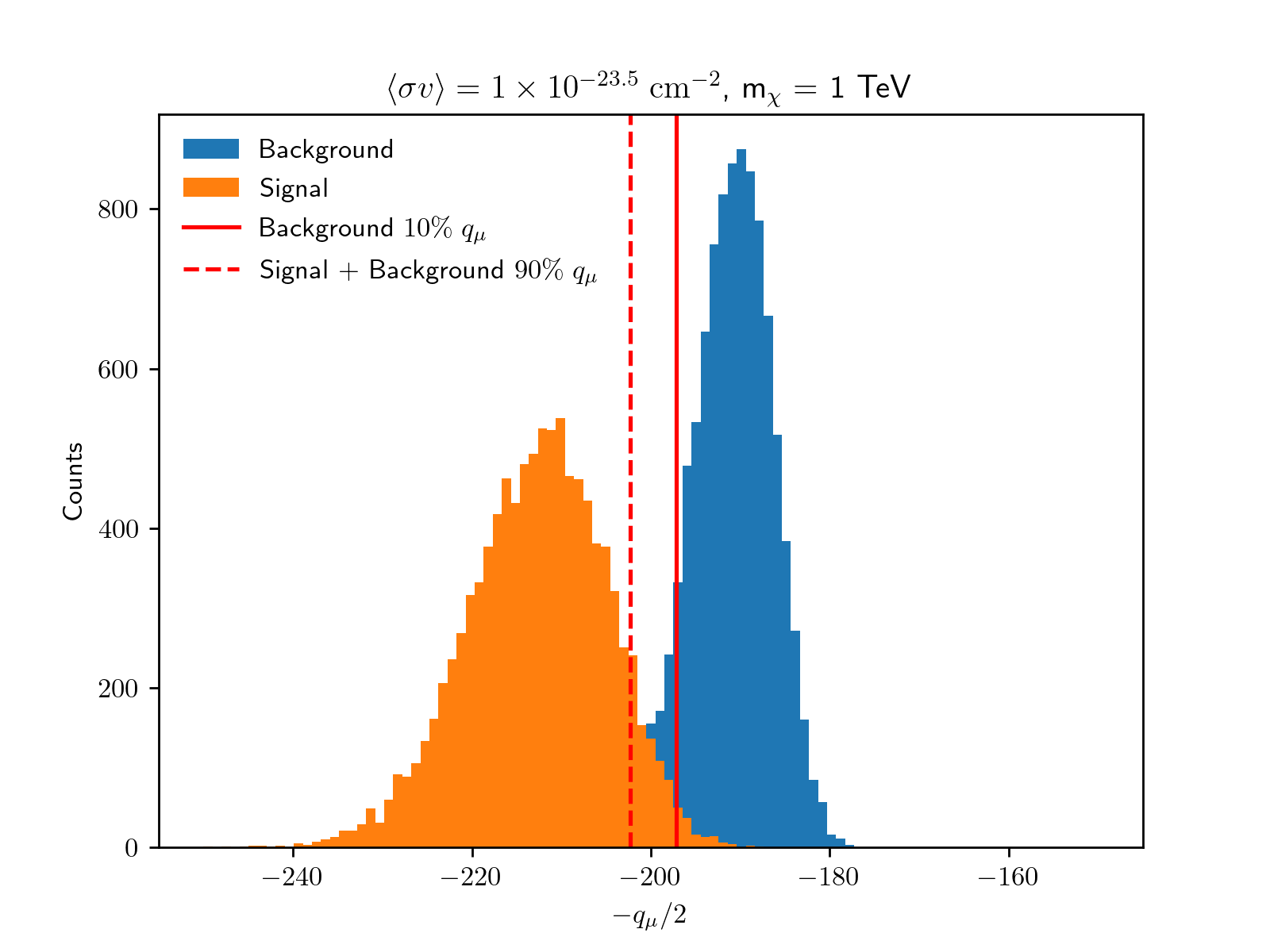}
    \caption{Example test statistic distributions for the background and signal hypothesis for IceCube. The red lines denote the 90\% (dashed) and 10\% (solid) quantiles for the signal and background hypothesis, respectively.}
    \label{fig:q_stat_ice}
    \end{center}
\end{figure}
\section{Confidence Level Example}\label{app:conf_ex}
\begin{figure}[H]
    \begin{center}
    \includegraphics[width=8.6cm]{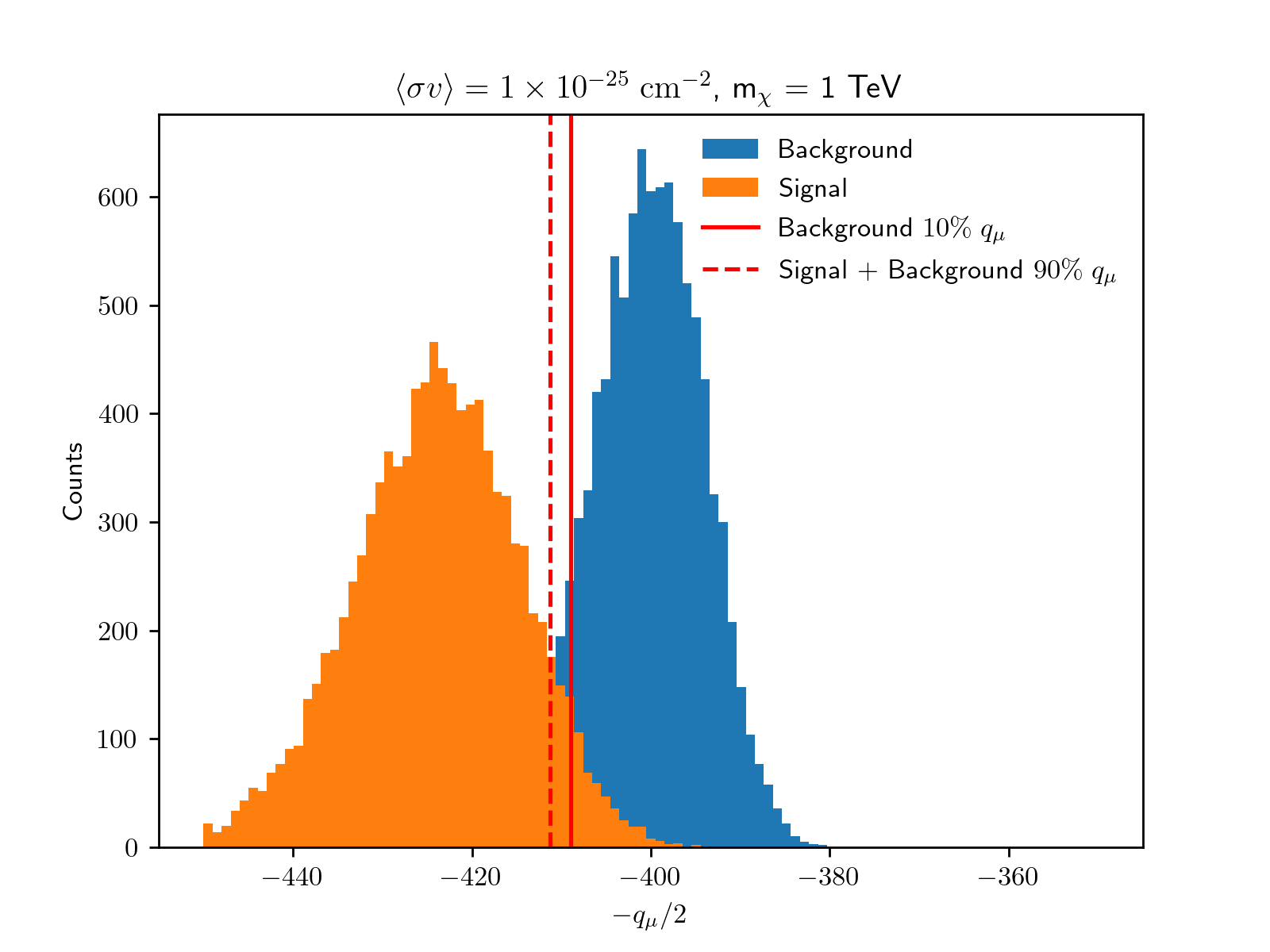}
    \caption{Example test statistic distributions for the background and signal hypothesis for P-ONE. The red lines denote the 90\% (dashed) and 10\% (solid) quantiles for the signal and background hypothesis, respectively.}
    \label{fig:q_stat}
    \end{center}
\end{figure}

In \autoref{fig:q_stat} and \autoref{fig:q_stat_ice}, we show an example distribution of the test statistic for P-ONE and IceCube.

\begin{figure}[H]
    \begin{center}
    \includegraphics[width=8.6cm]{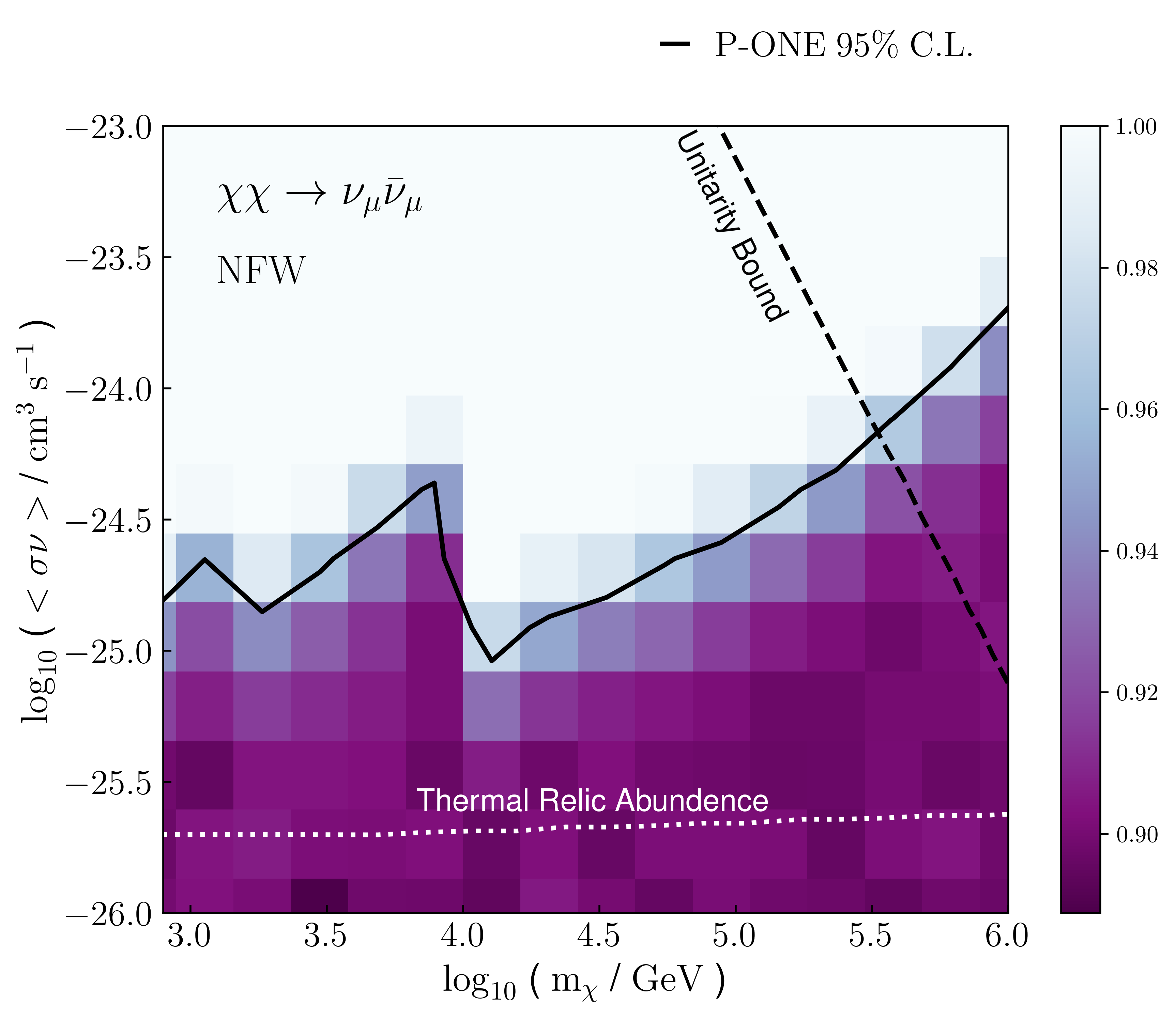}
    \caption{The confidence level plot for P-ONE. We have drawn the 95\% confidence level contour for reference.}
    \label{fig:pone_cmap}
    \end{center}
\end{figure}
\begin{figure}[h]
    \begin{center}
    \includegraphics[width=8.6cm]{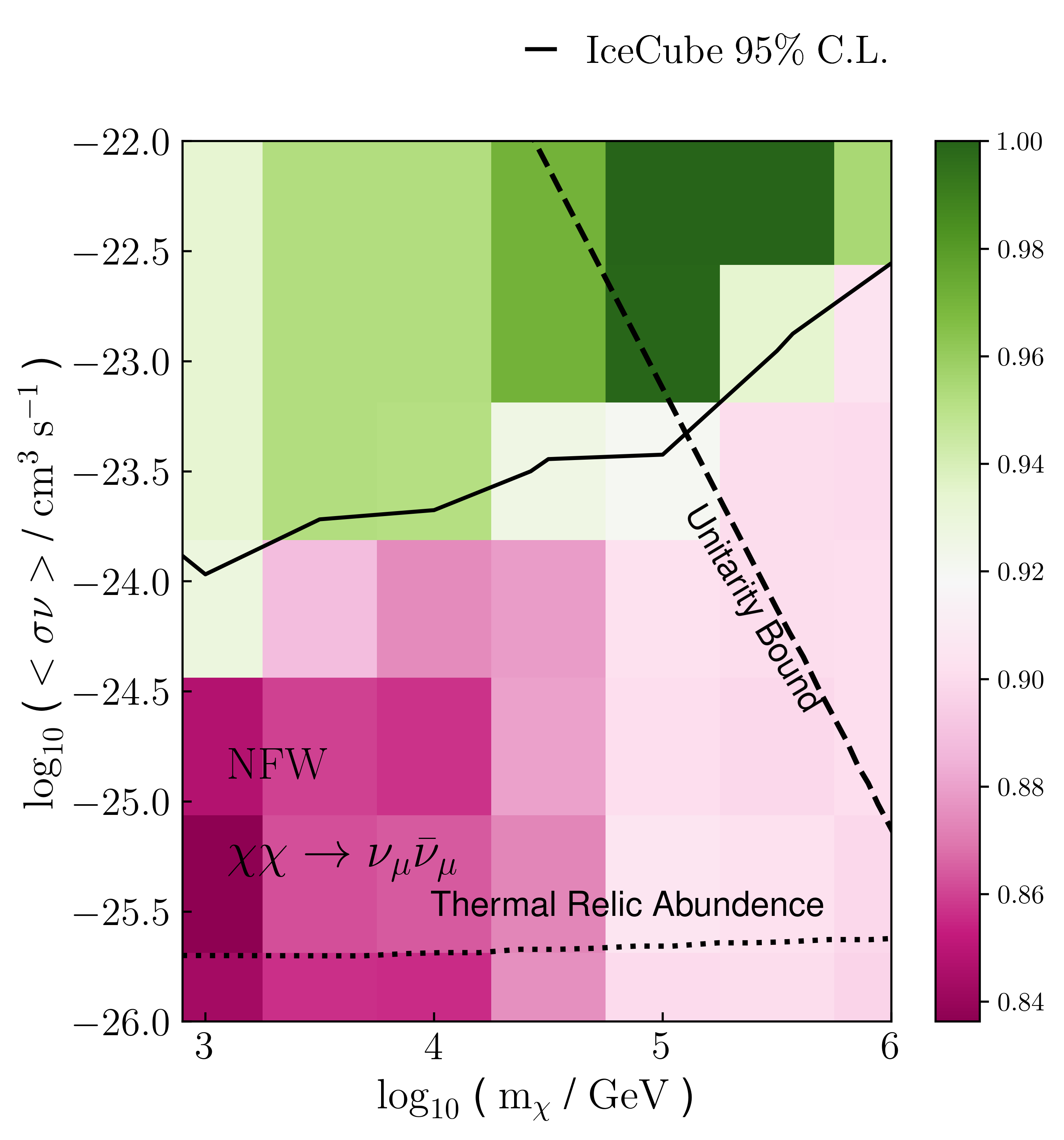}
    \caption{The confidence level plot  for IceCube. We have drawn the 95\% confidence level contour for reference.}
    \label{fig:cmap_ice}
    \end{center}
\end{figure}
 \autoref{fig:pone_cmap} and  \autoref{fig:cmap_ice}  show the 95 \% confidence levels (solid line) for P-ONE and IceCube respectively. This illustrates the distribution of confidence levels.

\section{Uncertainties due to atmospheric and astrophysical neutrino fluxes}\label{app:background_uncertainties}

\begin{figure}[h]
    \begin{center}
    \includegraphics[width=8.6cm]{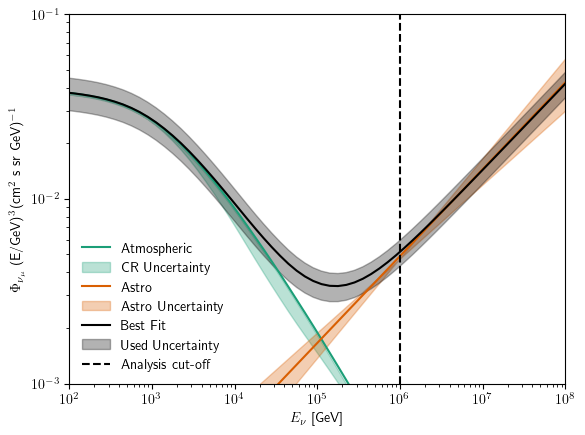}
    \caption{Here we show the uncertainties due to the cosmic ray flux (green) and the astrophysical neutrino flux (orange). We compare these uncertainties to the effects of $\eta_1$ and $\eta_2$ (gray) used in the analysis here.}
    \label{fig:bkgrd_uncertainties}
    \end{center}
\end{figure}

In \autoref{fig:bkgrd_uncertainties} we compare the uncertainties caused by CR and astrophysical neutrino uncertainties to the normalization uncertainties we introduce with $\eta_1$ and $\eta_2$. In the analysis range we consider here, CR and astrophysical uncertainties are smaller than those from $\eta_1$ and $\eta_2$. The uncertainties from the CR flux are calculated by injecting different primary cosmic ray models, from \cite{Gaisser:2013bla}.

\section{Analysis Tests}\label{app:analysis_test}

\begin{figure}[h]
    \begin{center}
    \includegraphics[width=8.6cm]{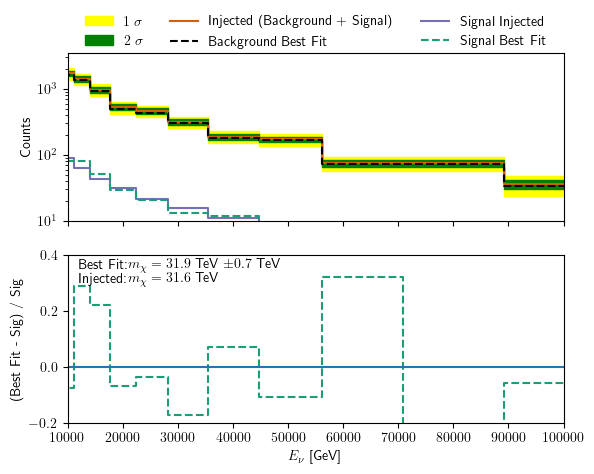}
    \caption{Comparing the injected signal and background (red) to the best-fit values for the background (black dashed) and signal (green dashed).}
    \label{fig:best_fit}
    \end{center}
\end{figure}

In this section, we test the analysis method employed here. In \autoref{fig:best_fit} we compare the injected counts with the obtained best-fit values.

\begin{figure}[h]
    \begin{center}
    \includegraphics[width=8.6cm]{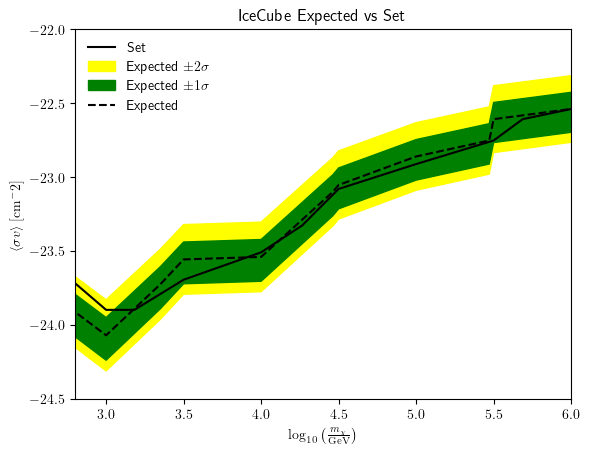}
    \caption{A comparison of the expected limit (dashed) to the one set using data (solid).}
    \label{fig:limits_exp_vs_measured}
    \end{center}
\end{figure}

In \autoref{fig:limits_exp_vs_measured}, we compare the expected limit to the one set using IceCube data.

\end{document}